\documentclass[12pt,onecolumn,journal,draftclsnofoot,a4paper]{IEEEtran}
\usepackage{cite}

\ifCLASSINFOpdf
\else
  \usepackage[dvips]{graphicx}
\fi
\usepackage{color}

\usepackage[cmex10]{amsmath}
\usepackage{amssymb}
\usepackage{multirow}
\usepackage{amsthm}
\allowdisplaybreaks[4]

\usepackage[tight,footnotesize]{subfigure}
\usepackage{url}
\usepackage{hyperref}
\hypersetup{
    bookmarksopen=true,
    bookmarksnumbered=true,
}

\newcommand{\ignore}[1]{}

\newtheorem{prop}{Proposition}
\newtheorem{corol}{Corollary}

\begin{document}
%
\title{Physical Layer Security in Cellular Networks: \\A Stochastic Geometry Approach}
%
%
%
\author{He~Wang,~\IEEEmembership{Student Member,~IEEE,}
        Xiangyun~Zhou,~\IEEEmembership{Member,~IEEE,}
        Mark~C.~Reed,~\IEEEmembership{Senior Member,~IEEE}
\thanks{This work was supported by National ICT Australia (NICTA), and the Australian Research Council¡¯s Discovery Projects funding scheme (Project No. DP110102548 and \ignore{Project No.} DP130101760). NICTA is funded by the Australian Government as represented by the Department of Broadband, Communications and the Digital Economy and the Australian Research Council through the ICT Centre of Excellence program.}
\thanks{H. Wang is with the Research School of Engineering, the Australian National University, ACT 0200, Australia, and also with Canberra Research Laboratory, National ICT Australia, ACT 2601, Australia (e-mail: he.wang@anu.edu.au).}
\thanks{X. Zhou is with the Research School of Engineering, the Australian National University, ACT 0200, Australia (e-mail: xiangyun.zhou@anu.edu.au).}
\thanks{M. C. Reed is with UNSW Canberra, ACT 2600, Australia, and also with the College of Engineering and Computer Science, the Australian National University, ACT 0200, Australia (e-mail: mark.reed@unsw.edu.au).}
}

\ifCLASSOPTIONdraftclsnofoot
\else
   \markboth{IEEE Transactions on Wireless Communications,~Vol.~xx, No.~xx, January~20xx}%
   {Wang \MakeLowercase{\textit{et al.}}: Physical Layer Security in Cellular Networks: A Stochastic Geometry Approach}
\fi
%



\maketitle

\begin{abstract}
  This paper studies the information-theoretic secrecy performance in large-scale cellular networks based on a stochastic geometry framework. The locations of both base stations and mobile users are modeled as independent two-dimensional Poisson point processes. We consider two important features of cellular networks, namely, information exchange between base stations and cell association, to characterize their impact on the achievable secrecy rate of an arbitrary downlink transmission with a certain portion of the mobile users acting as potential eavesdroppers. In particular, tractable results are presented under diverse assumptions on the availability of eavesdroppers' location information at the serving base station, which captures the benefit from the exchange of the location information between base stations.
\end{abstract}

\ifCLASSOPTIONdraftclsnofoot
\else
\fi

%
\IEEEpeerreviewmaketitle

\section{Introduction}\label{sec:Introduction}
%
%
%
%

During the past decades, we have witnessed the advancement of cellular communication networks\ignore{, which have become an important part of our daily life}. Because of the broadcast nature of the wireless medium, an unauthorized receiver located within the transmission range is capable of eavesdropping the unicast transmissions towards legitimate users, and security is always a crucial issue in cellular systems. Traditionally, most of security techniques in modern cellular standards, such as Wideband Code-Division Multiple Access (WCDMA) and Long Term Evolution (LTE), involve means of encryption algorithms in the upper layers of the protocol stacks \cite{3GPP_TS33.102,San09MCOM}. In contrast, the concept of achieving information-theoretic security by protecting the physical layer of wireless networks has attracted attention widely in the research community. Wyner proposed the wiretap channel model and the notion of perfect secrecy for point-to-point communication in his pioneering work \cite{Wyn75JBST}, which was extended to broadcast channels with confidential messages by Csisz{\'a}r and K{\"o}rner \cite{CsiKor78JIT}. Based on these initial results, a positive secrecy capacity, defined as the maximum transmission rate at which the eavesdropper is unable to obtain any information, can be achieved if the intended receiver enjoys a better channel than the potential eavesdropper.

Unlike point-to-point scenarios, the communication between nodes in large-scale networks strongly depends on the location distribution and the interactions between nodes. Based on the assumption that legitimate nodes and eavesdroppers are distributed randomly in the space, the studies on the secure communications for large-scale wireless networks have been carried out recently, from the information-theoretic viewpoint. Secrecy communication graphs describing secure connectivity over a large-scale network with eavesdroppers present were investigated in \cite{Hae08ISIT, PinBar12TIFS1, GoeAgg10ISIT, PinWin10ISITA}. In particular, the statistical characterizations of in-degree and out-degree under the security constraints were considered by Haenggi \cite{Hae08ISIT}, Pinto \emph{et al.} \cite{PinBar12TIFS1} and Goel \emph{et al.} \cite{GoeAgg10ISIT}. By using the tools from percolation theory, the existence of a secrecy graph was analyzed in \cite{Hae08ISIT, PinWin10ISITA}. The results in \cite{ZhoGan11JWCOM1} showed the improvements in the secure connectivity by introducing directional antenna elements and eigen-beamforming. In order to derive the network throughput, these works on connectivity were further extended for secrecy capacity analysis. Specifically, the maximum achievable secrecy rate under the worst-case scenario with colluding eavesdroppers was given in \cite{PinBar12TIFS2}. Scaling laws for secrecy capacity in large networks have been investigated in \cite{KoyKok12JIT, LiaPoo09ISIT, CapGoe12Infocom}. Focusing on the transmission capacity of secure communications, the throughput cost of achieving a certain level of security in an interference-limited network was analyzed in \cite{ZhoGan11JWCOM2, ZhoGan11Allerton}. It should be noticed that all works mentioned above were concentrated on ad hoc networks.

\subsection{Approach and Contributions}
In this work, we focus on the secrecy performance in large-scale cellular networks, considering cellular networks' unique characteristics different from ad hoc networks: the carrier-operated high-speed backhaul networks connecting individual base stations (BSs) and the core-network infrastructures, which provide us potential means of BS cooperation, such as associating mobile users to the optimal BS with secrecy considerations and exchanging information to guarantee better secure links.

Fortunately, modeling BSs to be randomly placed points in a plane and utilizing stochastic geometry \cite{StoKen95Book,BacBla09Book} to analyze cellular networks have been used extensively as an analytical tool for improving tractability \cite{Bro00JSAC, YanPet03TSP, Hae08TIT}. Recent works \cite{AndBac11JCOM, DhiGan12JSAC, WanQue12JSAC, CheNgu12VTC,YuKim11arXiv} have shown that the network models with BS locations drawn from a homogeneous Poisson point process (PPP) are as accurate as the traditional grid models compared with the result of an practical network deployment, and can provide more tractable analytical results which give pessimistic lower bounds on coverage and throughput. For these reasons we adopt PPPs to model the locations of BSs of the cellular networks in this paper.

The following scenario of secure communication in cellular networks is considered in this work: confidential messages are prepared to be conveyed to a mobile user, while certain other mobile users should not have the access to the messages and hence are treated as potential eavesdroppers. The serving BS should ensure the messages delivered to the intended user successfully while keeping perfect secrecy against all potential eavesdroppers. Considering the fact that the cellular service area is divided into cells, each BS knows the location as well as the identity of each user (i.e., whether the user is a potential eavesdropper or not) in its own cell. The identity and location information of mobile users in the other cells can be obtained by information exchange between BSs via the backhaul networks.

The main contributions of this paper are as follows:
\begin{itemize}
  \item First, our analytical results quantify the secrecy rate performance in large-scale cellular networks. Specifically, tractable results are provided on the probability distribution of the secrecy rate and hence the average secrecy rate achievable for a randomly located mobile user in such a cellular network, under different assumptions on the cell association and location information exchange between BSs as follows:
        \begin{itemize}
          \item \emph{Scenario-I}: the serving BS fully acquires potential eavesdroppers' location information; the nearest BS from the intended user is chosen as the serving BS.
          \item \emph{Scenario-II}: the serving BS fully acquires potential eavesdroppers' location information; the BS providing best secrecy performance at the intended user is chosen as the serving BS.
          \item \emph{Scenario-III}: the serving BS partially acquires potential eavesdroppers' location information; the nearest BS from the intended user is associated as the serving BS.
        \end{itemize}
  \item In addition, a unique feature of secure transmissions that the optimal BS is often not the nearest BS is identified and analyzed in the work. Our results show that only marginal gain can be obtained by optimally choosing the serving BS rather than associating to the nearest one. In other words, keeping the nearest BS to be used for secure transmission still achieves near-optimal secrecy performance, which is a very useful message to the network designers.
  \item Finally, our analysis sheds light into the impact of the availability of eavesdroppers' location information on the achievable secrecy rate. In particular, the secrecy performances for the scenarios with no location information exchange and limited exchange with neighboring cells are derived, which demonstrate the critical role of this kind of BS cooperation. This result provides network designers with practical guidelines in deciding on the necessary information exchange range, i.e., how many nearby BSs should participate in the information exchange for achieving a certain level of secrecy performance.
\end{itemize}

It should be noted that similar work to evaluate secrecy performance of large-scale cellular networks was conducted in \cite{SarHae10IM}; however, it mainly focused on the scaling behavior of the eavesdropper's density to allow full coverage over the entire network, without taking the achievable secrecy rate into account. In contrast, we characterize the statistics of the secrecy rate at an arbitrary mobile user under different cell association models and eavesdroppers' location information exchanging assumptions mentioned above.

The remainder of the paper is organized as follows: In Section~\ref{sec:SysModel}, we present the system model and general assumptions in this work. Section~\ref{sec:MainResults} shows the main result of this paper, in which we obtain simple tractable expressions for achievable secrecy rates under different scenarios. Section~\ref{sec:NumResults} provides numerical results and concluding remarks are given in Section~\ref{sec:Conclusion}.


\section{System Model}\label{sec:SysModel}
We consider the downlink scenario of a cellular network utilizing an orthogonal multiple access technique and composed of a single class of BSs, macro BS for instance. We focus on the performance achieved by a randomly chosen typical mobile user. The BSs are assumed to be spatially distributed as a two-dimensional homogeneous PPP $\Phi_{BS}$ of density $\lambda_{BS}$, and all BSs have the same transmit power value $P_{BS}$. An independent collection of mobile users, located according to an independent homogeneous PPP $\Phi_{MS}$ of density $\lambda_{MS}$, is assumed. We consider the process $\Phi_{MS} \cup \{0\}$ obtained by adding a user at the origin of the coordinate system. By Slivnyak's Theorem \cite{StoKen95Book}, this user can be taken as the typical user, since adding a user is identical to conditioning on a user at that location.
%

\subsection{Signal Model}\label{subsec:SignalModel}
The standard power loss propagation model is used with path loss exponent $\alpha > 2$\ignore{ and the complex fading coefficient is denoted by $h$. A special case where the wireless environment introduces only path loss, i.e., $h = 1$, is considered in this work}. Hence, the received power at the receiver $x_i$ from the transmitter $x_j$ is written as
  \begin{equation}\label{eqn:rxpower}
    P_{rx}(x_i,x_j) = P_{BS} \|{x_i-x_j}\|^{-\alpha}.
  \end{equation}
The noise power is assumed to be additive and constant with value $\sigma^2$ for all users, but no specific distribution is assumed.

In this work, we assume that there is no in-band interference at downlink receivers. This assumption is achievable by a carefully planned frequency reuse pattern, where the interfering BSs are far away to have the serving BS occupying some resource blocks exclusively in a relatively large region, and the interference can be incorporated in the constant noise power.

\subsection{Achievable Secrecy Rate}\label{subsec:SecCap}
We consider a scenario where confidential messages are prepared to be delivered to the typical user, while certain individuals among other mobile users, treated as potential malicious eavesdroppers (or called Eve for brevity) by the network, should be kept from accessing them. We model a fraction of the other mobile users randomly chosen from $\Phi_{MS}$ (the process constructed by all other users except the typical user) as the eavesdroppers, i.e., a thinned PPP with the density of $\lambda_e$, denoted by $\Phi_{e}$.

\begin{figure}[t!]
  \centering
  \includegraphics[width=0.54\textwidth, bb = 162 283 441 563, clip = true]{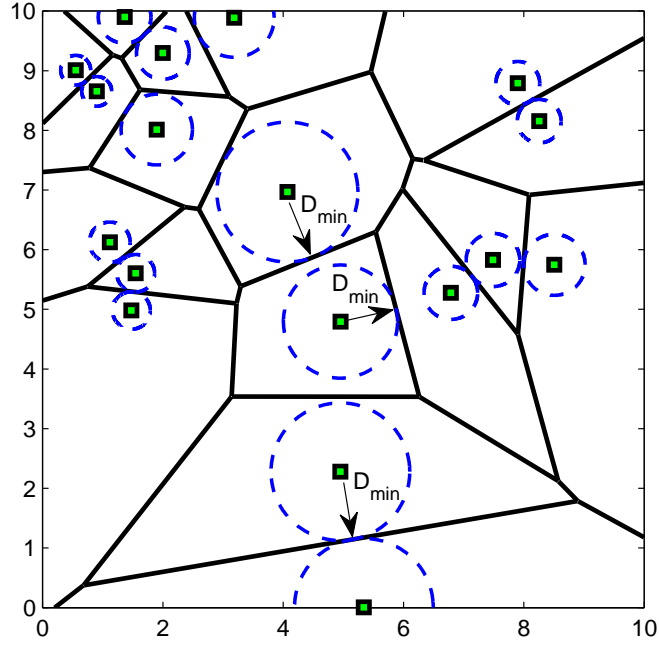}
  \caption{Illustration of Poisson distributed BSs' cell boundaries. Each user is associated with the nearest BS, and BSs (represented by green squares) are distributed according to PPP. $D_{\min}$ is defined as BS's minimum distance to its cell boundaries.}
  \label{fig:VT_Dmin} 
\end{figure}

Here we assume that each BS knows both the location and the identity (i.e., whether the user is a potential eavesdropper or not) of each mobile user in its own cell, and the cell of each BS is the Voronoi cell containing the BS, where the Voronoi tessellation is formed by PPP $\Phi_{BS}$ \cite{StoKen95Book}, as shown in Fig.~\ref{fig:VT_Dmin}. The identity and location information of mobile users in the other Voronoi cells can be obtained by the information exchange between BSs via backhaul networks.

Firstly, if we suppose the ideal case where the serving BS located at $x$ knows the locations of all eavesdroppers in the plane, which requires that the location and identity information of all users is shared completely through the backhaul network, the maximum secrecy rate achievable at the typical mobile user is given by \cite{PinBar12TIFS1, BloBar08JIT}, as
  \begin{equation}\label{eqn:secrecycapacity}
    R_s = \max \bigg\{\log_2 \Big(1+\frac{P_{rx}(0,x)}{\sigma^2} \Big) - 
    \log_2 \Big(1+\frac{P_{rx}(e^*(x),x)}{\sigma^2} \Big),0 \bigg\},
  \end{equation}
where
  \begin{equation}\label{eqn:e_star_fullinformation}
    e^*(x) = {\displaystyle \arg \max_{e \in {\Phi}_e}} P_{rx}(e,x) = {\displaystyle \arg \min_{e \in {\Phi}_e}} \|{e-x}\| ,
  \end{equation}
i.e., $e^*(x)$ is the location of the most detrimental eavesdropper, which is the nearest one from the serving BS in this case.

Then, assuming limited information exchange between BSs, there will be regions in which the eavesdroppers' location information is unknown to the serving BS, which is denoted by $\Theta \subset \mathbb{R}^2$. When this happens, the serving BS assumes the worst case, i.e., eavesdroppers can lie at any points in $\Theta$. Then the achievable secrecy rate is still given by (\ref{eqn:secrecycapacity}), but $e^*(x)$ should~be
  \begin{equation}\label{eqn:e_star}
    e^*(x) = {\displaystyle \arg \max_{e \in {\Phi}_e \cup \Theta}} P_{rx}(e,x),
  \end{equation}
where the detrimental eavesdropper is chosen from the union of the eavesdropper set ${\Phi}_e$ and the unknown region $\Theta$.

It should be noticed that the randomness introduced by $\Phi_{BS}$ and $\Phi_{e}$ makes the achievable secrecy rate $R_s$ at the typical user a random variable. Furthermore, the distribution of $R_s$ is mixed, i.e., $R_s$ has a continuous distribution on $(0,\infty)$ and a discrete component at $0$. For $R_s \in (0,\infty)$, the complementary cumulative distribution function (CCDF) of $R_s$ is given as
  \begin{equation}\label{eqn:Cs_CCDF_1}
    \bar{F}_{R_s}(R_0) =  \mathbb{P} \bigg(\log_2 \Big(1+\frac{P_{rx}(0,x)}{\sigma^2} \Big) -
    \log_2 \Big(1+\frac{P_{rx}(e^*(x),x)}{\sigma^2} \Big)>R_0 \bigg), \text{   for } R_0 \geqslant 0.
  \end{equation}
For the special case of $R_s = 0$, it has the probability $\mathbb{P}(R_s = 0) =  1 - \bar{F}_{R_s}(0)$, which corresponds to the probability that the link to the typical user cannot support any positive secrecy rate.

By assuming that the receivers of both the legitimate user and eavesdroppers operate in the high signal-to-noise ratio (SNR) regime, i.e., ${P_{rx}(0,x)}/{\sigma^2} \gg 1$ and ${P_{rx}(e^*(x),x)}/{\sigma^2} \gg 1$, we can obtain an approximation of $R_s$, denoted by $\hat{R}_s$, i.e., ${\hat{R}_s} = \max \big\{\log_2 \big({P_{rx}(0,x)}/{\sigma^2} \big) - \log_2 \big({P_{rx}(e^*(x),x)}/{\sigma^2} \big), 0\big\}$, the CCDF of which is
  \begin{eqnarray}\label{eqn:Cs_CCDF_2}
    \bar{F}_{\hat{R}_s}(R_0) & = & \mathbb{P} \Big(\frac{P_{rx}(0,x)}{P_{rx}(e^*(x),x)} > \beta \Big) \nonumber \\ 
    & = & \mathbb{P} \Big( \|{e^*(x)-x}\| > \beta^{1/\alpha} \|{x}\| \Big), \text{   for } R_0 \geqslant 0,
  \end{eqnarray}
where the threshold $\beta$ is defined as $\beta \triangleq 2^{R_0}$.
In this work, we focus on high SNR scenarios and use the above expression to obtain tractable results on the secrecy rate performance. The obtained analytical results give approximations on the secrecy performance at finite SNR values.

Furthermore, from the fact that the achievable secrecy rate $R_s$ should always be non-negative, we can easily reach the conclusion that the high SNR approximation $\bar{F}_{\hat{R}_s}(R_0)$ serves as an upper bound for the CCDF of $R_s$ at finite SNR, i.e.,
  \begin{eqnarray}\label{eqn:Cs_CCDF_3}
      \bar{F}_{R_s}(R_0) & = & \mathbb{P} \Big(\frac{\sigma^2 + P_{rx}(0,x)}{\sigma^2 + P_{rx}(e^*(x),x)} > 2^{R_0} \Big) \nonumber \\
      & \leqslant & \mathbb{P} \Big(\frac{P_{rx}(0,x)}{P_{rx}(e^*(x),x)} > 2^{R_0} \Big) = \bar{F}_{\hat{R}_s}(R_0), \text{  for } R_0 \geqslant 0,
  \end{eqnarray}
where the two probability expressions are equal when $R_0 = 0$.
Therefore, some of our analytical results on $\bar{F}_{\hat{R}_s}(R_0)$ and $\mathbb{E}[\hat{R}_s]$ under the high SNR assumption, including the exact expressions and upper bounds, give valid upper bounds on the secrecy performances at finite SNR values.

\section{Main Results}\label{sec:MainResults}
In this section, we provide the main results on the probabilistic characteristics of the achievable secrecy rates $\hat{R}_s$ and the average secrecy rates achievable $\mathbb{E}[\hat{R}_s]$ under three major scenarios, where different criterions to choose the serving BS are used and the serving BS can fully or partially acquire the location information of the eavesdroppers, corresponding to the different levels of BS cooperation introduced. It should be noticed that the BS cooperation considered in this paper includes only exchanging the identity and location information of the mobile users and selecting the appropriate BS to serve the typical user.

\subsection{Scenario-I: Full Location Information; Nearest BS to Serve}\label{subsec:FullLocInfo_NearestBS}
We firstly assume that the location information of all eavesdroppers can be fully accessed by the serving BS and employ the cell association model by confining mobile users to be served by the nearest BS only. The location and identity information of mobile users in the serving BS's cell can be obtained easily, and other users' information is supplied by other BSs via the backhaul networks. Associating users to the nearest BS is commonly used in related cellular modeling works \cite{Bro00JSAC,AndBac11JCOM}, and equivalently it means that a BS is associated with the users in its Voronoi cell (formed by the PPP $\Phi_{BS}$). \ignore{, thus resulting the Voronoi tessellation for BS coverage areas, as shown in Fig.~\ref{fig:VT_Dmin}.}

\begin{prop} \label{prop:FullLocInfo_NearestBS}
Under the conditions of mobile users being served by the nearest BS and the availability of full location information for all eavesdroppers, the CCDF of the achievable secrecy rate obtained at the typical user is given by
  \begin{eqnarray}\label{eqn:Cs_CCDF_scn-I}
    \bar{F}_{\hat{R}_s}(R_0) = \frac{1}{1+\frac{\lambda_e}{\lambda_{BS}} \cdot 2^{(2R_0)/\alpha} }, \text{   for } R_0 \geqslant 0.
  \end{eqnarray}
\end{prop}

\begin{IEEEproof}
Here we use $x_0$ to denote the nearest BS from the origin, and we define $r_u$ as the distance from the typical user to the nearest BS, namely, $r_u = \|{x_0}\|$. The probability density function (pdf) of $r_u$ has been provided in \cite{Hae05TIT}, as
  \begin{align}\label{eqn:r_u_pdf}
    f_{r_u}(r) = 2 \pi \lambda_{BS} r \exp(-\pi \lambda_{BS} r^2).
  \end{align}
Due to the assumption that the serving BS knows all eavesdroppers' locations in this scenario, the most detrimental eavesdropper $e^*(x_0)$ for the BS at $x_0$ should be the nearest one from $x_0$, as given in (\ref{eqn:e_star_fullinformation}). We define the (closed) ball centered at $p$ and of radius $r$ as $\mathcal{B}(p,r)$, i.e., $\mathcal{B}(p,r) \triangleq \{m \in \mathbb{R}^2, \|m-p\| \leqslant r\}$. Then the CCDF of the achievable secrecy rate $\hat{R}_s$ under this scenario can be derived as
   \begin{eqnarray}\label{eqn:Cs_CCDF_derive_scn-I}
    \bar{F}_{\hat{R}_s}(R_0) & = & \mathbb{P} \Big( \|{e^*(x_0)-x_0}\| > \beta^{1/\alpha} \|{x_0}\| \Big) \nonumber \\
    & = & \int_0^\infty \mathbb{P} \Big( \|{e^*(x_0)-x_0}\| > \beta^{1/\alpha} r_u \mid r_u = y \Big) f_{r_u}(y) \mathrm{d} y \nonumber \\
    & = & \int_0^\infty \mathbb{P} \big[ \text{No Eve in } \mathcal{B}(x_0,\beta^{1/\alpha} r_u)\mid r_u = y \big] f_{r_u}(y) \mathrm{d} y \nonumber \\
    & \stackrel{(a)}{=} & \int_0^\infty \mathbb{P} \big[ \text{No Eve in } \mathcal{B}(x_0,\beta^{1/\alpha} y) \big] f_{r_u}(y) \mathrm{d} y \nonumber \\
    & \stackrel{(b)}{=} & \int_0^\infty \exp(-\pi \lambda_e \beta^{2/\alpha} y^2) \cdot 2 \pi \lambda_{BS} y \exp(-\pi \lambda_{BS} y^2) \mathrm{d} y \nonumber \\
    & = & \frac{1}{1+\frac{\lambda_e}{\lambda_{BS}} \cdot 2^{(2R_0)/\alpha} },
  \end{eqnarray}
where step $(a)$ is derived based on the independence between $\Phi_e$ and $\Phi_{BS}$, and step $(b)$ follows the PPP's void probability and pdf of $r_u$ given in (\ref{eqn:r_u_pdf}). Through the deduction above, the CCDF expression of the achievable secrecy rate can be obtained.
\end{IEEEproof}

\begin{corol}\label{corol:FullLocInfo_NearestBS}
Under the conditions of mobile users being served by the nearest BS and the availability of full location information for all eavesdroppers, the average secrecy rate achievable at the typical user is provided by
  \begin{eqnarray}\label{eqn:Cs_Avg_scn-I}
    \mathbb{E}[\hat{R}_s] = \frac{\alpha}{2 \ln2} \cdot \ln \Big(\frac{\lambda_{BS} + \lambda_{e}}{\lambda_{e}}\Big).
  \end{eqnarray}
\end{corol}

\begin{IEEEproof}
Based on the CCDF expression given in Proposition~\ref{prop:FullLocInfo_NearestBS}, the average secrecy rate achievable at the typical user can be obtained by integrating (\ref{eqn:Cs_CCDF_scn-I}) from $0$ to $\infty$, i.e.,
  \begin{eqnarray}\label{eqn:Cs_Avg_derive_scn-I}
      \mathbb{E}[\hat{R}_s]  & = & \int_0^{\infty} \frac{1}{1+\frac{\lambda_e}{\lambda_{BS}} \cdot 2^{(2 t)/\alpha} } \mathrm{d} t \nonumber \\
      & \stackrel{(a)}{=} & \bigg[\frac{1}{\ln(2^{2/\alpha})}\cdot \ln \bigg(\frac{\exp\big[\ln(2^{2/\alpha})t\big]}{1+\frac{\lambda_e}{\lambda_{BS}}\cdot \exp\big[\ln(2^{2/\alpha})t\big]}\bigg) \bigg]_0^\infty \nonumber \\
      & = & \frac{\alpha}{2 \ln2} \ln\Big(\frac{1}{{\lambda_e}/{\lambda_{BS}}}\Big) - \frac{\alpha}{2 \ln2} \ln\Big(\frac{1}{1+{\lambda_e}/{\lambda_{BS}}}\Big) \nonumber \\
      & = & \frac{\alpha}{2 \ln2} \cdot \ln \Big(\frac{\lambda_{BS} + \lambda_{e}}{\lambda_{e}}\Big),
  \end{eqnarray}
where step $(a)$ follows the indefinite integral result for the form of the integrand herein, which can be found in \cite{JefDai08HandBook}.
\end{IEEEproof}

\subsection{Scenario-II: Full Location Information; Optimal BS to Serve}\label{subsec:FullLocInfo_BestBS}
Next, we still keep the assumption that the serving BS has all eavesdroppers' location information, which can be achieved by an ideal information exchange between BSs; however, in this scenario, we assume that all BSs can act as candidates to serve the typical user.

This scenario provides us the maximum achievable secrecy rate from the information-theoretic point of view, which tells the network designer the ultimate secrecy performance the cellular network can offer and can be viewed as the optimal BS cooperation scheme considered in this paper. Obviously, to obtain the optimal secrecy performance, the BS achieving the maximum secrecy rate should be selected. By studying the secrecy performance with the optimal cell association, we are able to quantify the gap between the secrecy performances provided by the optimal BS and the nearest BS.

Based upon these assumptions, the achievable secrecy rate at the typical user becomes
  \begin{eqnarray}\label{eqn:secrecycapacity_scn-II}
    \hat{R}_s = \max \bigg\{\max_{x \in \Phi_{BS}} \Big\{\log_2 \Big(\frac{P_{rx}(0,x)}{\sigma^2} \Big) -
    \log_2 \Big(\frac{P_{rx}(e^*(x),x)}{\sigma^2} \Big)\Big\},0 \bigg\},
  \end{eqnarray}
where $e^*(x)$ is given by (\ref{eqn:e_star_fullinformation}).

\begin{prop} \label{prop:FullLocInfo_BestBS}
Under the conditions of mobile users being served by the optimal BS and the availability of full location information for all eavesdroppers, an upper bound for the CCDF of the achievable secrecy rate at the typical user is given by
  \begin{align}\label{eqn:Cs_upperCCDF_scn-II}
    \bar{F}_{\hat{R}_s}(R_0) \leqslant 1 - \exp \Big( - \frac{\lambda_{BS}}{\lambda_e 2^{(2 R_0)/\alpha}} \Big), \text{   for } R_0 \geqslant 0,
  \end{align}
and a lower bound is given by
  \begin{align}\label{eqn:Cs_lowerCCDF_scn-II}
    \bar{F}_{\hat{R}_s}(R_0) \geqslant \frac{1}{1+\frac{\lambda_e}{\lambda_{BS}} \cdot 2^{(2R_0)/\alpha} }, \text{   for } R_0 \geqslant 0.
  \end{align}
\end{prop}

\begin{IEEEproof}
For a given BS (not necessarily the nearest BS) located at the position of $x$, its achievable secrecy rate toward the origin's typical user is larger than $R_0$ if and only if there is no eavesdroppers located within $\mathcal{B}(x, 2^{(R_0/\alpha)} \|x\|)$. Hence, the achievable secrecy rate's cumulative distribution function (CDF) can be derived as
  \begin{eqnarray}\label{eqn:Cs_CDF_derive_scn-II}
    {F}_{\hat{R}_s}(R_0) & = & \mathbb{P} ( \hat{R}_s \leqslant R_0 ) \nonumber \\
    & = & \mathbb{P} \big[\text{All BSs can not provide secrecy rate larger than } R_0\big] \nonumber \\
    & = & \mathbb{E}_{\Phi_{e}} \bigg[  \mathbb{E}_{\Phi_{BS}} \Big[ \prod_{x\in\phi_{BS}} \mathbf{1}\big\{\Phi_e \bigcap \mathcal{B}(x, 2^{\frac{R_0}{\alpha}} \|x\|) \neq 0 \big\} \Big] \bigg] \nonumber \\
    & = & \mathbb{E}_{\Phi_{e}} \Bigg[  \mathbb{E}_{\Phi_{BS}} \bigg[ \prod_{x\in\phi_{BS}} \Big[ 1 - \mathbf{1}\big\{\Phi_e \bigcap \mathcal{B}(x, 2^{\frac{R_0}{\alpha}} \|x\|) = 0 \big\} \Big] \bigg] \Bigg] \nonumber \\
    & \stackrel{(a)}{=} & \mathbb{E}_{\Phi_{e}} \bigg[  \exp \Big[- \lambda_{BS} \int_{\mathbb{R}^2} \mathbf{1}\big\{\Phi_e \bigcap \mathcal{B}(x, 2^{\frac{R_0}{\alpha}} \|x\|) = 0 \big\} \mathrm{d} x \Big] \bigg] \nonumber \\
    & \geqslant & \exp \bigg[- \lambda_{BS} \int_{\mathbb{R}^2} \mathbb{P}\Big[\Phi_e\big(\mathcal{B}(x, 2^{\frac{R_0}{\alpha}} \|x\|)\big) = 0 \Big]  \mathrm{d} x \bigg],
  \end{eqnarray}
where $R_0 \geqslant 0$, step $(a)$ follows from the probability generating functional (PGFL) of the PPP \cite{StoKen95Book}, and Jensen's inequality gives the lower bound for ${F}_{\hat{R}_s}(R_0)$ in the last step. The part in the integral can be derived by using 2-D homogeneous PPP's void probability \cite{StoKen95Book}, i.e., $\mathbb{P}\big[\Phi_e\left(\mathcal{B}(x, 2^{(R_0/\alpha)} \|x\|)\right) = 0 \big] = \exp(-\pi \lambda_e 2^{(2 R_0/\alpha)} \|x\|^2)$, which can be substituted into the integration in (\ref{eqn:Cs_CDF_derive_scn-II}) to obtain the upper bound of the achievable secrecy rate's CCDF in (\ref{eqn:Cs_upperCCDF_scn-II}) easily.

Then we turn to find the lower bound for the CCDF of the achievable secrecy rate.
Here we use $\hat{R}_{s, nearest}$ to denote the achievable secrecy rate where only the nearest BS is accessible, which has been studied in Scenario-I.
Since connecting to the nearest BS is always one of the viable options if all BSs are reachable, we can have the usual stochastic order between $\hat{R}_{s, nearest}$ in Scenario-I and $\hat{R}_s$ in the current scenario, i.e., $\mathbb{P}(\hat{R}_{s, nearest} > R_0) \leqslant \mathbb{P}(\hat{R}_s > R_0)$ or equivalently $\bar{F}_{\hat{R}_s}(R_0) \geqslant \bar{F}_{\hat{R}_{s, nearest}}(R_0)$. Therefore, the conclusion in Proposition~\ref{prop:FullLocInfo_NearestBS} provides the lower bound in (\ref{eqn:Cs_lowerCCDF_scn-II}), which completes the proof.
\end{IEEEproof}

\begin{prop} \label{prop:FullLocInfo_BestBS_2nd_UpperBound}
Under the conditions of mobile users being served by the optimal BS and the availability of full location information for all eavesdroppers, another upper bound for the CCDF of the achievable secrecy rate at the typical user is given by
  \begin{eqnarray}\label{eqn:Cs_upperCCDF_2nd_scn-II}
    \bar{F}_{\hat{R}_s}(R_0) \leqslant 1 - \mathbb{E}_{V_d} \Big[ \exp \big( - \frac{4}{(1+ 2^{R_0/\alpha})^2} \cdot \frac{\lambda_{BS}}{\lambda_e} \cdot V_d \big) \Big], \text{   for } R_0 \geqslant 0,
  \end{eqnarray}
where the expectation is taken over the random variable $V_d$, the area of the typical Voronoi cell of a PPP with the unitary density.
\end{prop}

\begin{IEEEproof}
For the set of eavesdropper locations $\Phi_{e}$, we can define a random set $\mathcal{P}$, the union of all points at which BS can provide the typical user (at the origin) a secrecy rate $\hat{R}_s > R_0$,~i.e.,
  \begin{align}\label{eqn:randomset_A_define}
    \mathcal{P}  {=}  \Big\{x \in \mathbb{R}^2: \|{e-x}\| > \beta^{1/\alpha} \|{x}\|, \forall e \in \Phi_{e} \Big\},
  \end{align}
which is based upon the assumption that the serving BS knows all eavesdroppers' locations in this scenario. Furthermore, we define $\mathcal{C}$ as the Voronoi cell generated by the process $\Phi_{e} \cup \{0\}$, the union of the eavesdroppers' locations and the origin. Because of Slivnyak's Theorem, the Voronoi cell around the origin formed by $\Phi_{e} \cup \{0\}$ has the same property as a randomly chosen Voronoi cell formed by a PPP with density $\lambda_e$. The area measures of the random set $\mathcal{P}$ and $\mathcal{C}$ are denoted by $A(\mathcal{P})$ and $A(\mathcal{C})$ respectively. An example of these random sets is illustrated in Fig.~\ref{fig:IrregularVoronoiTessellation}, in which we can obtain a straightforward relationship between $A(\mathcal{P})$ and $A(\mathcal{C})$ as
  \begin{eqnarray}\label{eqn:Area_Relationship}
    A(\mathcal{P})  \leqslant \frac{4}{(1+\beta^{1/\alpha})^2} A(\mathcal{C}),
  \end{eqnarray}
if $\beta \geqslant 1$ or equivalently $R_0 \geqslant 0$.

The value $\left[{4}/{(1+\beta^{1/\alpha})^2}\right] A(\mathcal{C})$ is the area measure of the region enclosed by blue lines in Fig.~\ref{fig:IrregularVoronoiTessellation}, which is the exact shape shrunk from $\mathcal{C}$ and has edges tangential to $\mathcal{P}$'s edges. Obviously, for a realization of the BS location $\Phi_{BS}$, the typical user can have a secrecy rate larger than $R_0$ if and only if there is at least a BS located in $\mathcal{P}$, which makes the CCDF of the secrecy rate $\hat{R}_s$ become
  \begin{eqnarray}\label{eqn:Cs_upperCCDF_2nd_derive_scn-II}
    \bar{F}_{\hat{R}_s}(R_0)
    & = & \mathbb{P} \big[\text{No BS exists in } \mathcal{P} \big] \nonumber \\
    & \stackrel{(a)}{=} & 1 - \mathbb{E}_{\mathcal{P}}\Big[\exp\big(-\lambda_{BS}A(\mathcal{P})\big)\Big] \nonumber \\
    & \leqslant & 1 - \mathbb{E}_{\mathcal{P}}\Big[\exp\big(- \frac{4 \lambda_{BS}}{(1+\beta^{1/\alpha})^2} A(\mathcal{C})\big)\Big] \nonumber \\
    & = & 1 - \mathbb{E}_{V_d}\Big[\exp\big(- \frac{4}{(1+\beta^{1/\alpha})^2} \cdot \frac{\lambda_{BS}}{\lambda_{e}} \cdot  {V_d}\big)\Big],
  \end{eqnarray}
where the expectation in step $(a)$ is taken over the random set~$\mathcal{P}$.
\begin{figure}[t!]
  \centering
  \includegraphics[width=0.54\textwidth, bb = 110 237 496 613, clip = true]{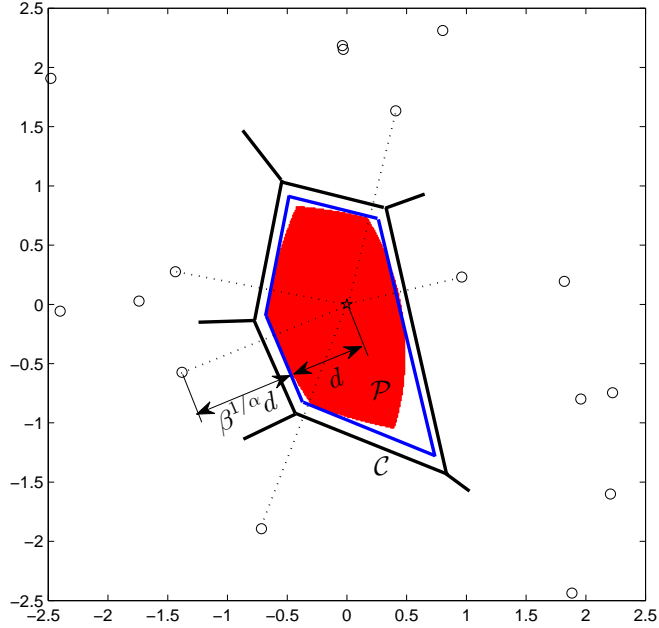}
  \caption{Illustration of the relationship between $\mathcal{P}$ (the union of all points at which BS can provide the typical user a secrecy rate $\hat{R}_s > \log_2(\beta)$, where $\beta = 1.25$, represented as the red region) and $\mathcal{C}$ (the Voronoi cell generated by the process $\Phi_{e} \cup \{0\}$), as defined in the proof of Proposition~\ref{prop:FullLocInfo_BestBS_2nd_UpperBound}. The typical user denoted by a star is located at the origin. A realization of eavesdroppers are scattered and denoted as circles. }
  \label{fig:IrregularVoronoiTessellation}
\end{figure}
\end{IEEEproof}

\textit{Remark:} It can be observed that the upper bound obtained in Proposition~\ref{prop:FullLocInfo_BestBS_2nd_UpperBound} depends on the statistic characteristics of Voronoi cell's area. It provides us an accurate approximation for small positive $\hat{R}_s$ values and complements the upper bound result in Proposition~\ref{prop:FullLocInfo_BestBS}. Particularly, for the special case of $R_0 = 0$, the region $\mathcal{P}$ turns out to be the Voronoi cell $\mathcal{C}$, thus making the CCDF upper bound become the exact result, i.e.,
  \begin{eqnarray}\label{eqn:Cs_CCDF_Rs0_scn-II}
    \bar{F}_{\hat{R}_s}(0) = 1 - \mathbb{E}_{V_d}\Big[\exp\big(- \frac{\lambda_{BS}}{\lambda_{e}}  {V_d}\big)\Big],
  \end{eqnarray}
and the expression in this extreme case is consistent with the secrecy coverage probability provided in \cite{SarHae10IM}.
For high value of $R_0$, however, the area difference between $A(\mathcal{P})$ and $\left[{4}/{(1+\beta^{1/\alpha})^2}\right] A(\mathcal{C})$ increases, which makes the approximation in (\ref{eqn:Area_Relationship}) become imprecise. This can explain the numerical results we will observe later in Fig.~\ref{fig:topic2_SecureConnectionOutage_C0C5}, i.e., the discrepancy between the upper bound given by Proposition~\ref{prop:FullLocInfo_BestBS_2nd_UpperBound} and the simulation result for $R_0 = 5$.

Although there is no known closed form expression of $V_d$'s pdf \cite{OkaBoo92Book}, some accurate estimates of this distribution were produced in \cite{HinMil80Math,WeaKer86Math}. For instance, a simple gamma distribution was used to fit the pdf of $V_d$ derived from Monte Carlo simulations in \cite{WeaKer86Math}, i.e.,
  \begin{align}\label{eqn:Vd_pdf}
    f_{V_d}(x) \approx b^q x^{q-1} \exp(-bx)/\Gamma(q),
  \end{align}
where $q = 3.61$, $b = 3.61$ and $\Gamma(x) = \int_0^{\infty} t^{x-1} e^{-t} \mathrm{d} t$ is the standard gamma function. By substituting this estimate into (\ref{eqn:Cs_upperCCDF_2nd_scn-II}) and simplifying the integral, we can obtain
  \begin{eqnarray}\label{eqn:Cs_upperCCDF_2nd_approx_scn-II}
    \bar{F}_{\hat{R}_s}(R_0) \approx 1 - \frac{b^q}{\big(b + \frac{4}{(1+ 2^{R_0/\alpha})^2} \cdot \frac{\lambda_{BS}}{\lambda_e} \big)^q}, 
    \text{   for } R_0 \geqslant 0.
  \end{eqnarray}

After giving the bounds for $\hat{R}_s$'s CCDF, we will focus on the average secrecy rate achievable for a randomly located user.

\begin{corol}\label{corol:FullLocInfo_BestBS}
Under the conditions of mobile users being served by the optimal BS and the availability of full location information for all eavesdroppers, an upper bound of the average secrecy rate achievable at the typical user is provided by
  \begin{align}\label{eqn:Cs_upperAvg_scn-II}
    \mathbb{E}[\hat{R}_s]
    & \leqslant \frac{\alpha}{2 \ln2} \cdot \Big[\gamma + \ln\big(\frac{\lambda_{BS}}{\lambda_{e}}\big) + E_1\big(\frac{\lambda_{BS}}{\lambda_{e}}\big) \Big],
  \end{align}
and a lower bound is provided by
  \begin{align}\label{eqn:Cs_lowerAvg_scn-II}
    \mathbb{E}[\hat{R}_s]  \geqslant \frac{\alpha}{2 \ln2} \cdot \ln \Big(\frac{\lambda_{BS} + \lambda_{e}}{\lambda_{e}}\Big),
  \end{align}
where $E_1(x) = \int_x^\infty \exp(-t) \frac{1}{t} \mathrm{d} t$ is the exponential integral and $\gamma$ is the Euler-Mascheroni constant.
\end{corol}

\begin{IEEEproof}
Based on the CCDF bounds given in Proposition~\ref{prop:FullLocInfo_BestBS}, the upper and lower bound of the average secrecy rate achievable at the typical user can be obtained by integrating (\ref{eqn:Cs_upperCCDF_scn-II}) and (\ref{eqn:Cs_lowerCCDF_scn-II}) from $0$ to $\infty$. Specifically, the upper bound can be derived as
  \begin{eqnarray}\label{eqn:Cs_upperAvg_derive_scn-II}
    \mathbb{E}[\hat{R}_s]
    & \leqslant & \int_0^{\infty} \Big[ 1 - \exp \Big( - \frac{\lambda_{BS}}{\lambda_e 2^{(2 t)/\alpha}} \Big) \Big] \mathrm{d} t \nonumber \\
    & \stackrel{(a)}{=} & \frac{1}{\ln(2^{2/\alpha})} \int_{0}^{\frac{\lambda_{BS}}{\lambda_e}} \frac{1 - \exp ( - v ) }{v} \mathrm{d} v,
  \end{eqnarray}
where step $(a)$ is derived by employing a change of variables $v = {\lambda_{BS}}/{(\lambda_e 2^{(2 t)/\alpha})}$. We use the Taylor series expansion of $\exp (-v)$, and the integrand in (\ref{eqn:Cs_upperAvg_derive_scn-II}) becomes
  \begin{eqnarray}\label{eqn:Cs_upperAvg_derive_2_scn-II}
     \frac{1 - \exp ( - v )}{v} = \sum_{k=1}^{\infty}\frac{(-v)^{k-1}}{k!}.
  \end{eqnarray}
Then by integrating both sides of the equation (\ref{eqn:Cs_upperAvg_derive_2_scn-II}) and performing simple mathematical operations, we can obtain the relationship
  \begin{eqnarray}\label{eqn:Cs_upperAvg_derive_3_scn-II}
     \int_0^{\frac{\lambda_{BS}}{\lambda_e}} \frac{1 - \exp ( - v )}{v}  \mathrm{d} v
     & = & \int_0^{\frac{\lambda_{BS}}{\lambda_e}} \sum_{k=1}^{\infty}\frac{(-v)^{k-1}}{k!} \mathrm{d} v \nonumber \\
     & = & \sum_{k=1}^{\infty} \int_0^{\frac{\lambda_{BS}}{\lambda_e}} \frac{(-v)^{k-1}}{k!} \mathrm{d} v \nonumber \\
     & = & - \sum_{k=1}^{\infty} \frac{(-\frac{\lambda_{BS}}{\lambda_e})^{k}}{k \cdot k!}.
  \end{eqnarray}
Since the exponential integral can be expressed as $E_1(x) = - \gamma - \ln(x) + \sum_{k=1}^{\infty} \frac{(-1)^{k+1} x^{k}}{k \cdot k!} $ when $x > 0$ \cite{AbrSte65HandBook}, the above integral can be derived as
  \begin{eqnarray}\label{eqn:Cs_upperAvg_derive_4_scn-II}
     \int_0^{\frac{\lambda_{BS}}{\lambda_e}} \frac{1 - \exp ( - v )}{v}  \mathrm{d} v
     = & \gamma + \ln\big(\frac{\lambda_{BS}}{\lambda_{e}}\big) + E_1\big(\frac{\lambda_{BS}}{\lambda_{e}}\big).
  \end{eqnarray}
Plugging (\ref{eqn:Cs_upperAvg_derive_4_scn-II}) into (\ref{eqn:Cs_upperAvg_derive_scn-II}) gives the upper bound of the average secrecy rate in (\ref{eqn:Cs_upperAvg_scn-II}).

On the other hand, following the same procedure as the one to prove Corollary~\ref{corol:FullLocInfo_NearestBS}, the lower bound of average secrecy rate can be obtained, which completes the proof.
\end{IEEEproof}

An alternative upper bound of the average secrecy rate achievable can be derived based upon Proposition~\ref{prop:FullLocInfo_BestBS_2nd_UpperBound}, and the corresponding performance will also be shown in Section~\ref{sec:NumResults}.

\begin{figure}[t!]
  \centering
  \includegraphics[width=0.64\textwidth, bb = 96 185 460 475, clip = true]{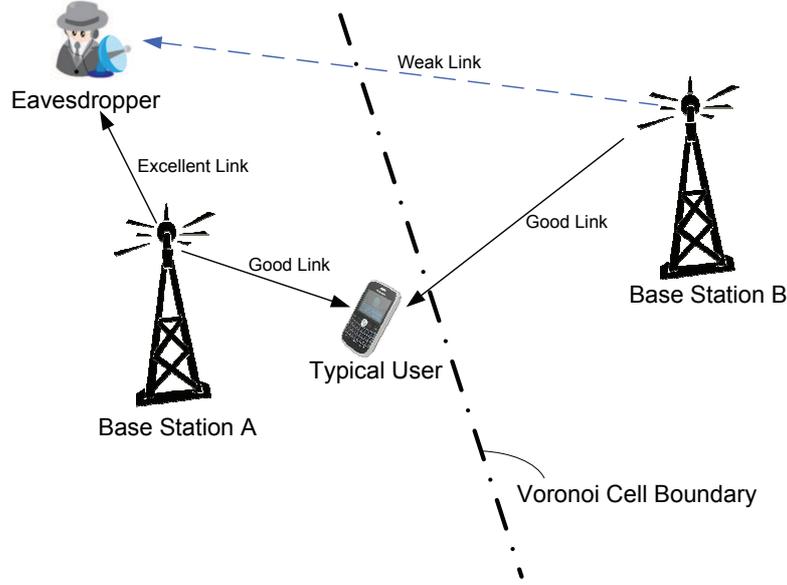}
  \caption{An example where the BS providing maximum achievable secrecy rate is not the nearest BS. The typical user's nearest BS is BS-A, which however cannot provide a positive secrecy rate due to its excellent link to the eavesdropper. BS-B, on the other hand, can provide a secrecy connection since there is no eavesdroppers nearby.}
  \label{fig:Most_Proper_BS}
\end{figure}

It should be noticed that the optimal BS mentioned here is not necessarily the nearest BS, since it is possible that other BSs can provide higher secrecy rate than the nearest BS. Taking the case illustrated in Fig.~\ref{fig:Most_Proper_BS} for example, the typical user's nearest BS is BS-A, which, however, is hardly capable of providing a secure connection due to its excellent connection to the eavesdropper nearby. Alternatively, choosing BS-B to serve can provide a certain level secrecy rate if the typical user's channel quality to BS-B is better than the channel to the eavesdropper.

By comparing the secrecy performance in Scenario-I (the typical user served by the nearest BS) with this scenario (the typical user served by the best BS), we will be able to see the benefit from optimally choosing the serving BS to provide the secure downlink transmission. The numerical illustrations will be provided in Section~\ref{sec:NumResults}.

\subsection{Scenario-III: Limited Location Information; Nearest BS to Serve}\label{subsec:NonFullLocInfo_NearestBS}
Here we still assume the same cell association model as Scenario-I, i.e., mobile users are served by the nearest BS, nevertheless only limited users' location and identity information is known to the serving BS. Considering the backhaul bandwidth cost in practice and the core-network implementation complexity for BS cooperation, the scenarios where the location and identity information is only exchanged with neighboring cells or even no exchange allowed at all are analyzed in this section.

\subsubsection{No location and identity information exchange}\label{subsubsec:NonInfoEx}
Firstly, we assume that no location and identity information exchange allowed between BSs, which means that the serving BS only knows the intracell users' location and identity information. As mentioned in section~\ref{subsec:SecCap}, the unknown region outside the serving cell leads to the worst case assumption that eavesdroppers lie on the serving BS's cell boundaries and limit the achievable secrecy rate.

Before coming to this scenario's secrecy performance, we firstly define the minimum distance from PPP's each point to its own cell boundaries, denoted as $D_{\min}$. In Fig.~\ref{fig:VT_Dmin}, for instant, the $D_{\min}$ of three BSs are illustrated. In the cell tessellation formed by BS PPP with density $\lambda_{BS}$, we can simply use the void probability of a PPP to derive
  \begin{align}\label{eqn:D_min_cdf}
    \mathbb{P}\left(D_{\min} > r \right) & = \mathbb{P}\big[\text{No BS closer than }2r\big] \nonumber \\
    & = e^{-\pi \lambda_{BS} (2r)^2}.
  \end{align}
Therefore, the CDF is $F_{D_{min}}(r) = \mathbb{P}\left(D_{\min} \leqslant r \right)  = 1 - e^{-\pi \lambda_{BS} (2r)^2}$ and the pdf can be found~as
  \begin{align}\label{eqn:D_min_pdf}
    f_{D_{\min}}(r) = \frac{\mathrm{d}F_{D_{min}}(r)}{\mathrm{d}r} = 8 \pi \lambda_{BS} r \exp(-4 \pi \lambda_{BS} r^2).
  \end{align}

\begin{prop} \label{prop:NonInfoEx}
Under the conditions of mobile users being served by the nearest BS and only intracell eavesdroppers' location information available, a lower bound for the CCDF of the achievable secrecy rate obtained at the typical user is given by
  \begin{align}\label{eqn:Cs_lowerCCDF_scn-IIIA}
    \bar{F}_{\hat{R}_s}(R_0) \geqslant \frac{1}{1+(\frac{\lambda_e}{\lambda_{BS}} + 4)\cdot 2^{(2R_0)/\alpha} }, \text{   for } R_0 \geqslant 0.
  \end{align}
\end{prop}

\begin{IEEEproof}
Based on the available intracell eavesdroppers' location information and the assumption that the typical user is served by the nearest BS at $x_0$, (\ref{eqn:Cs_CCDF_2}) becomes
  \begin{eqnarray}\label{eqn:Cs_CCDF_derive_scn-IIIA}
    \bar{F}_{\hat{R}_s}(R_0) & = & \mathbb{P} \Big( \|{e^*(x_0)-x_0}\| > \beta^{1/\alpha} \|{x_0}\| \Big) \nonumber \\
    & \stackrel{(a)}{=} & \mathbb{P} \big[ \text{No Eve in } \mathcal{B}(x_0,\beta^{\frac{1}{\alpha}} r_u) ; r_u < \beta^{-\frac{1}{\alpha}}D_{\min} \big] \nonumber \\
    & = & \int_{0}^{\infty} \mathbb{P} \big[ \text{No Eve in } \mathcal{B}(x_0,\beta^{\frac{1}{\alpha}} r_u); D_{\min} > \beta^{\frac{1}{\alpha}} r_u \mid r_u = y \big] f_{r_u}(y) \mathrm{d} y,
  \end{eqnarray}
where step $(a)$ is based on the fact that eavesdroppers are assumed to be lied in the cell boundaries for the worst case. The probability expression herein can be further derived~as
  \begin{eqnarray}\label{eqn:Prob_derive_scn-IIIA}
    & & \mathbb{P} \big[ \text{No Eve in } \mathcal{B}(x_0,\beta^{\frac{1}{\alpha}} r_u); D_{\min} > \beta^{\frac{1}{\alpha}} r_u  \mid r_u = y \big] \nonumber \\
    & \stackrel{(b)}{=} & \mathbb{P} \big[ \text{No Eve in } \mathcal{B}(x_0,\beta^{\frac{1}{\alpha}} y) \big] \cdot \mathbb{P} \big( D_{\min} > \beta^{\frac{1}{\alpha}} y  \mid r_u = y \big) \nonumber \\
    & \geqslant & \mathbb{P} \big[ \text{No Eve in } \mathcal{B}(x_0,\beta^{\frac{1}{\alpha}} y) \big] \cdot \mathbb{P} \big( D_{\min} > \beta^{\frac{1}{\alpha}} y  \big) \nonumber \\
    & = & \exp \big(-\pi \lambda_e (\beta^{\frac{1}{\alpha}} y)^2 \big) \int_{\beta^{\frac{1}{\alpha}} y}^{\infty}  f_{D_{\min}}(z) \mathrm{d} z \nonumber \\
    & = & \exp \big(-\pi (\lambda_e + 4 \lambda_{BS}) \beta^{\frac{2}{\alpha}} y^2 \big),
  \end{eqnarray}
where the independence between $\Phi_e$ and $\Phi_{BS}$ is used to separate the two probability expressions in step $(b)$, and the former part is only dependent on the density of eavesdroppers $\lambda_e$ and the ball's area $\pi \beta^{2/\alpha}y^2$, but independent of $x_0$. It should be noticed that the value of $r_u$ has an impact on the distribution of $D_{min}$, and we need to use $f_{D_{\min} \mid r_u}(\cdot \, | \, \cdot)$ to derive $\mathbb{P} ( D_{\min} > \beta^{\frac{1}{\alpha}} y  \mid r_u = y )$ in step $(b)$. Because the tractable result of $f_{D_{\min} \mid r_u}(\cdot \, | \, \cdot)$ is not available, we obtain a lower bound (also served as a tractable approximation) expression by ignoring the impact of $r_u$ on the distribution of $D_{min}$, due to the fact that $\mathbb{P} ( D_{\min} > x  \mid r_u = y ) \geqslant \mathbb{P} \left( D_{\min} > x \right)$. The lower bound by replacing distribution $f_{D_{min}}(\cdot)$ can provide a good approximation, which will be demonstrated by the numerical comparisons in Section~\ref{sec:NumResults}.

By substituting (\ref{eqn:Prob_derive_scn-IIIA}) and the pdf of $r_u$ given in (\ref{eqn:r_u_pdf}) into (\ref{eqn:Cs_CCDF_derive_scn-IIIA}), the lower bound expression (\ref{eqn:Cs_lowerCCDF_scn-IIIA}) can be obtained, which completes the proof.
\end{IEEEproof}

\textit{Remark:} When $\lambda_e \gg \lambda_{BS}$, the impact of cell boundaries on the secrecy rate becomes negligible, since almost surely an eavesdropper exists inside the ball $\mathcal{B}(x_0, D_{min})$ and limits the achievable secrecy rate, then making (\ref{eqn:Cs_lowerCCDF_scn-IIIA}) become (\ref{eqn:Cs_CCDF_scn-I}).

\begin{corol}\label{corol:NonInfoEx}
Under the conditions of mobile users being served by the nearest BS and only intracell eavesdroppers' location information available, a lower bound of the average secrecy rate achievable at the typical user is provided by
  \begin{eqnarray}\label{eqn:Cs_lowerAvg_scn-IIIA}
    \mathbb{E}[\hat{R}_s] \geqslant \frac{\alpha}{2 \ln2} \cdot \ln \Big(\frac{5 \lambda_{BS} + \lambda_{e}}{4 \lambda_{BS} + \lambda_{e}}\Big).
  \end{eqnarray}
\end{corol}
\begin{IEEEproof}
The lower bound of the average secrecy rate $\mathbb{E}[\hat{R}_s]$ can be derived by integrating (\ref{eqn:Cs_lowerCCDF_scn-IIIA}) from $0$ to $\infty$. Since the integrand in this integral has the similar form as (\ref{eqn:Cs_CCDF_scn-I}), the same deduction procedure can be performed to obtain this lower bound.
\end{IEEEproof}

\textit{Remark:} Under the condition of mobile users camping on the nearest BS, Scenario-I and this case can be regarded as two extremes: in the former scenario, the location information of all eavesdroppers is shared among BSs, while no location and identity information exchange is allowed in the latter one. By comparing the expressions of (\ref{eqn:Cs_Avg_scn-I}) with (\ref{eqn:Cs_lowerAvg_scn-IIIA}), it is easy to conclude that the latter case's average secrecy rate achievable increases with $\lambda_{BS}/\lambda_e$ much slower than the counterpart in Scenario-I. This trend, which will be given numerically in following Section~\ref{sec:NumResults}, demonstrates the impact of the location and identity information exchange between BSs.


\subsubsection{Location and identity information exchange limited with neighboring cells only}\label{subsubsec:detectionD}
In order to further characterize how the availability of the location and identity information affects the secrecy performance, we will investigate the secrecy rate for the case where the location information and identity exchange is restricted to the serving BS's neighboring cells only.

Given certain neighboring BSs participating in the information exchange with the serving BS, the region outside the cells covered by these BSs is the unknown region. By considering the worst case scenario that the eavesdroppers can be located anywhere inside the unknown region, the secrecy performance is limited by the minimum distance from the unknown region to the serving BS. As long as the minimum distance is the same, the secrecy performance stays the same regardless of the shape of the unknown region, which means that the consideration of a disk-shape known region does not lose the generality of the result on secrecy rates. Therefore, we apply the following model to represent the known and unknown regions: only the location information of the eavesdroppers with distances less than $D_0$ from the serving BS is available to it, i.e., the eavesdroppers outside the region $\mathcal{B}(x,D_0)$ are unknown to a BS at $x$. The value $D_0$ is called \emph{detection radius} in our analysis.

From a network design perspective, a larger $D_0$ represents information exchanging feasible with BSs farther away, and in other words, a larger $D_0$ means that more BSs participate in the information exchange with the serving BS. This scenario provides limited information exchange, which can be regarded as an intermediate case between Scenario-II and Scenario-III(1), and reflects practical considerations, such as the limited bandwidth of the backhaul network and the complexity introduced by extensive information sharing in the practical implementation. By investigating how the achievable secrecy rate changes with $D_0$, one can obtain insights on the improvement of the secrecy performance as more BSs participate in the information exchange process.

\begin{prop} \label{prop:detectionD}
Under the conditions of mobile users being served by the nearest BS and the detection radius is $D_0$, the CCDF of the achievable secrecy rate obtained at the typical user is given by
  \begin{eqnarray}\label{eqn:Cs_CCDF_scn-IIIB}
    \bar{F}_{\hat{R}_s}(R_0) = \Big(1- \exp\big[-\pi (\lambda_e+\lambda_{BS} 2^{-\frac{2R_0}{\alpha}}) {D_0}^2\big]\Big) 
    \cdot \frac{1}{1+\frac{\lambda_e}{\lambda_{BS}} \cdot 2^{(2R_0)/\alpha} }, \ \ \text{   for } R_0 \geqslant 0.
  \end{eqnarray}
\end{prop}

\begin{IEEEproof}
Based on the available location information of eavesdroppers with distances less than $D_0$ and the typical user served by the nearest BS at $x_0$, (\ref{eqn:Cs_CCDF_2}) can be derived as
  \begin{eqnarray}\label{eqn:Cs_CCDF_derive_scn-IIIB}
    \bar{F}_{\hat{R}_s}(R_0) & = & \mathbb{P} \Big( \|{e^*(x_0)-x_0}\| > \beta^{1/\alpha} \|{x_0}\| \Big) \nonumber \\
    & = & \mathbb{P} \big[ \text{No Eve in } \mathcal{B}(x_0,\beta^{\frac{1}{\alpha}} r_u) ; r_u < \beta^{-\frac{1}{\alpha}}D_{0} \big] \nonumber \\
    & = & \int_{0}^{\beta^{-\frac{1}{\alpha}} D_{0}} \mathbb{P} \big[ \text{No Eve in } \mathcal{B}(x_0,\beta^{\frac{1}{\alpha}} r_u) \mid r_u = y \big] f_{r_u}(y) \mathrm{d} y \nonumber \\
    & \stackrel{(a)}{=} & \int_{0}^{\beta^{-\frac{1}{\alpha}} D_{0}} \mathbb{P} \big[ \text{No Eve in } \mathcal{B}(x_0,\beta^{\frac{1}{\alpha}} y) \big] f_{r_u}(y) \mathrm{d} y \nonumber \\
    & \stackrel{(b)}{=} & \int_{0}^{2^{-\frac{R_0}{\alpha}} D_{0}} 2 \pi \lambda_{BS} y \cdot \exp(-\pi \lambda_e 2^{\frac{2 R_0}{\alpha}} y^2 - \pi \lambda_{BS} y^2) \mathrm{d} y \nonumber \\
    & = & \frac{1}{1+\frac{\lambda_e}{\lambda_{BS}} \cdot 2^{(2R_0)/\alpha} } \cdot \Big(1- \exp\big[-\pi (\lambda_e+\lambda_{BS} 2^{-\frac{2R_0}{\alpha}}) {D_0}^2\big]\Big),
  \end{eqnarray}
where step $(a)$ follows the independence between $\Phi_e$ and $\Phi_{BS}$, and step $(b)$ is derived based on the void probability of PPP and the pdf of $r_u$. It should be noticed that the probability expression $\mathbb{P} \big[ \text{No Eve in } \mathcal{B}(x_0,\beta^{\frac{1}{\alpha}} y) \big]$ is only dependent on the density of eavesdroppers $\lambda_e$ and the ball's area $\pi \beta^{2/\alpha}y^2$, but independent of $x_0$. The integration from $0$ to $2^{-\frac{R_0}{\alpha}} D_{0}$ gives the result which completes the proof.
\end{IEEEproof}

\textit{Remark:} As expected, the general trend can be understood as follows: when detection radius $D_0$ decreases, the location information of eavesdroppers surrounding the serving BS reduces, which makes a lower probability to maintain the secrecy rate $R_0$. As we increase $D_0$ to infinity, the condition turns to be the same as Scenario-I, thus making (\ref{eqn:Cs_CCDF_scn-IIIB}) become~(\ref{eqn:Cs_CCDF_scn-I}).

\begin{corol}\label{corol:detectionD}
Under the conditions of mobile users being served by the nearest BS and the detection radius is $D_0$, the average secrecy rate achievable at the typical user is provided~by
  \begin{eqnarray}\label{eqn:Cs_Avg_scn-IIIB}
    \mathbb{E}[\hat{R}_s] = \frac{\alpha}{2 \ln2} \cdot \ln \Big(\frac{\lambda_{BS} + \lambda_{e}}{\lambda_{e}}\Big) 
    - \frac{\alpha}{2 \ln2} \cdot \Big[E_1\big(\pi \lambda_e D_0^2\big) - E_1\big(\pi (\lambda_e + \lambda_{BS}) D_0^2\big)\Big].
  \end{eqnarray}
\end{corol}
\begin{IEEEproof}
Based on the CCDF expression given in Proposition~\ref{prop:detectionD}, the average secrecy rate achievable at the typical user can be provided by integrating (\ref{eqn:Cs_CCDF_scn-IIIB}) from $0$ to $\infty$, i.e.,
  \begin{eqnarray}\label{eqn:Cs_Avg_derive_scn-IIIB}
    \mathbb{E}[\hat{R}_s]
    & = & \int_0^{\infty} \frac{1}{1+\frac{\lambda_e}{\lambda_{BS}} \cdot 2^{(2t)/\alpha} } \cdot \Big(1- \exp\big[-\pi (\lambda_e+\lambda_{BS} 2^{-\frac{2t}{\alpha}}) D_0^2\big]\Big) \mathrm{d} t \nonumber \\
    & = & \int_0^{\infty} \frac{1}{1+\frac{\lambda_e}{\lambda_{BS}} \cdot 2^{(2t)/\alpha} }\mathrm{d} t - \int_0^{\infty} \frac{\exp\big[-\pi (\lambda_e+\lambda_{BS} 2^{-\frac{2t}{\alpha}}) D_0^2\big]}{1+\frac{\lambda_e}{\lambda_{BS}} \cdot 2^{(2t)/\alpha} }\mathrm{d} t \nonumber \\
    & \stackrel{(a)}{=} & \frac{\alpha}{2 \ln2} \cdot \ln \Big(\frac{\lambda_{BS} + \lambda_{e}}{\lambda_{e}}\Big) - \nonumber \\
    & & \ \ \ \ \ \ \ \ \exp(-\pi \lambda_e D_0^2) \int_0^{\infty} \frac{\exp\big[-\pi \lambda_{BS} D_0^2 \cdot 2^{-\frac{2t}{\alpha}}\big]}{1+\frac{\lambda_e}{\lambda_{BS}} \cdot 2^{(2t)/\alpha} }\mathrm{d} t,
  \end{eqnarray}
where the deduction of the former part in step $(a)$ utilizes the result solved in Corollary~\ref{corol:FullLocInfo_NearestBS}, and then we will focus on the integral in its latter part, i.e.,
  \begin{eqnarray}\label{eqn:Cs_Avg_derive_2_scn-IIIB}
     & & \exp(-\pi \lambda_e D_0^2) \int_0^{\infty} \frac{\exp\big[-\pi \lambda_{BS} D_0^2 \cdot 2^{-\frac{2t}{\alpha}}\big]}{1+\frac{\lambda_e}{\lambda_{BS}} \cdot 2^{(2t)/\alpha} }\mathrm{d} t \nonumber \\
     & \stackrel{(b)}{=} & \exp(-\pi \lambda_e D_0^2) \int_{\frac{\lambda_e}{\lambda_{BS}}}^{\infty} \frac{\exp(-\pi \lambda_e D_0^2 v^{-1})}{1 + v} \cdot \frac{\alpha}{2 v \ln2 } \mathrm{d} v \nonumber \\
     & \stackrel{(c)}{=} & \frac{\alpha}{2 \ln2} \int_{\pi \lambda_e D_0^2}^{\pi (\lambda_{BS} + \lambda_e) D_0^2} \frac{1}{s \exp(s)} \mathrm{d} s \nonumber \\
     & = & \frac{\alpha}{2 \ln2} \Big[E_1\big(\pi \lambda_e D_0^2\big) - E_1\big(\pi (\lambda_e + \lambda_{BS}) D_0^2\big)\Big],
  \end{eqnarray}
where step $(b)$ and step $(c)$ are obtained by employing changes of variables $v = \frac{\lambda_e}{\lambda_{BS}} \cdot 2^{(2t)/\alpha}$ and $s = \frac{\pi \lambda_e D_0^2}{v} + \pi \lambda_e D_0^2$ respectively, and the last step can be derived by using the definition of the exponential integral. Plugging (\ref{eqn:Cs_Avg_derive_2_scn-IIIB}) into (\ref{eqn:Cs_Avg_derive_scn-IIIB}) gives the desired result in (\ref{eqn:Cs_Avg_scn-IIIB}), which completes the proof.
\end{IEEEproof}

\section{Numerical Illustrations}\label{sec:NumResults}
In this section, we present numerical results on the achievable secrecy rate for all three major scenarios respectively. Here we define the value $\mathrm{{SNR}}$ as the received SNR from the serving BS at the distance $r=1$, i.e., $\mathrm{SNR} = {P_{BS}}/{\sigma^2}$. All simulation results are conducted under a high SNR condition, i.e., $\mathrm{SNR} = 20 \mathrm{dB}$, and unitary BS density, i.e., $\lambda_{BS} = 1$, to compare with our analysis for the purpose of model validation.

\begin{figure}[t!]
  \centering
  \includegraphics[width=0.64\textwidth, bb = 110 260 478 575, clip = true]{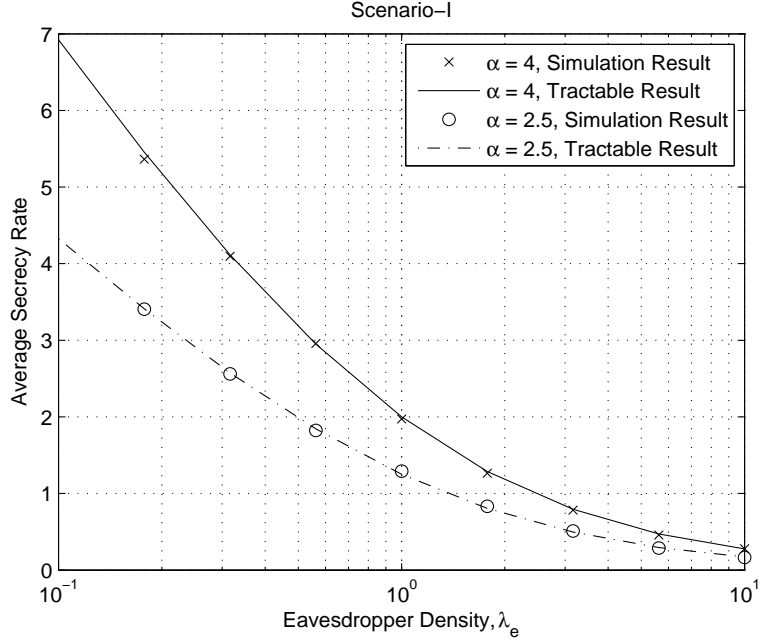}
  \caption{The average secrecy rate achievable versus the eavesdropper density $\lambda_e$ for
           Scenario-I (full location information; nearest BS to serve).
           Simulation and tractable analytical results are shown for different path loss exponents $\alpha$.}
  \label{fig:topic1_AverageSecureRate} 
\end{figure}

Firstly, for each curve in Fig.~\ref{fig:topic1_AverageSecureRate}, we show the average secrecy rates achievable at the typical user in Scenario-I, for both path loss exponents of $\alpha = 4$ and $\alpha = 2.5$. As can be seen in this figure, the curves representing the analytical expression (\ref{eqn:Cs_Avg_scn-I}) in Corollary~\ref{corol:FullLocInfo_NearestBS} match the simulated results for all conditions.

\begin{figure}[t]
  \centering
  \subfigure{
    \label{fig:topic2_SecureConnectionOutage_C0C5:a} 
    \includegraphics[height=0.275\textheight, bb = 98 253 488 585, clip = true]{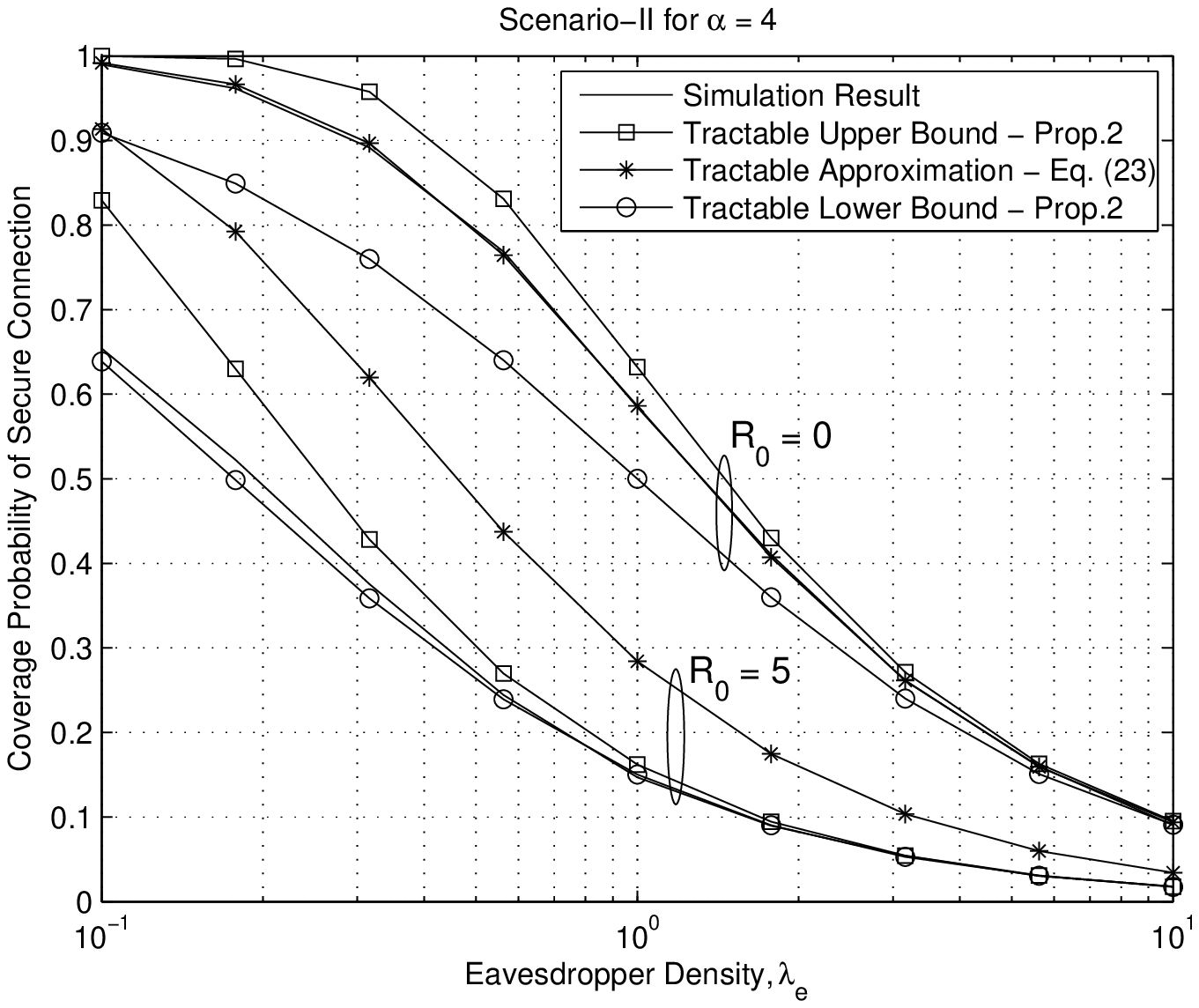}}
  \subfigure{
    \label{fig:topic2_SecureConnectionOutage_C0C5:b} 
    \includegraphics[height=0.275\textheight, bb = 98 253 488 585, clip = true]{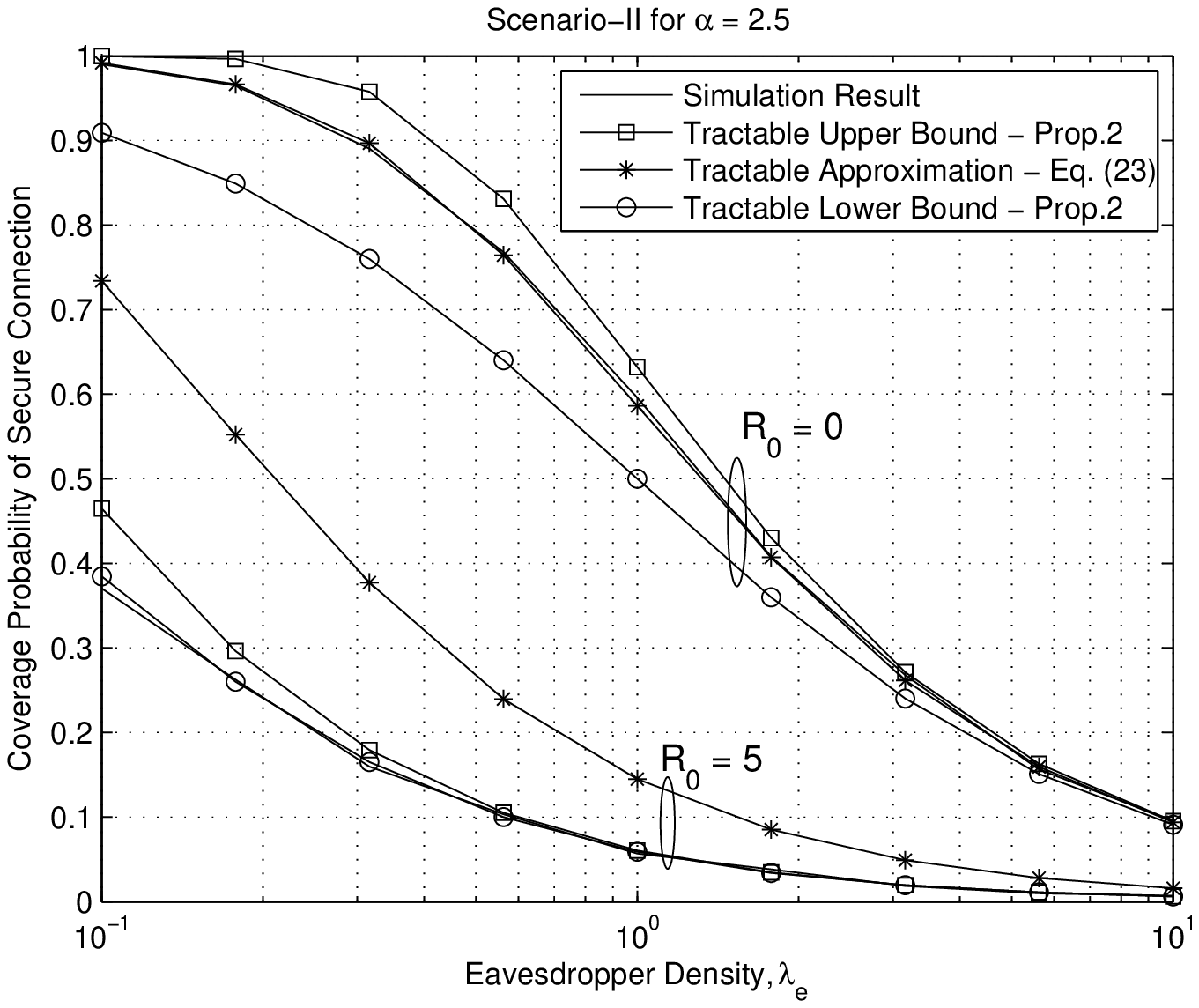}}
  \caption{The secure coverage probability versus the eavesdropper density $\lambda_e$ for
           Scenario-II (full location information; optimal BS to serve).
           Simulation and tractable analytical results are shown for different thresholds $R_0 = 0$ or $5$ to claim outage.
           \ignore{(It should be noticed that the approximation provided by Proposition~\ref{prop:FullLocInfo_BestBS_2nd_UpperBound} becomes the exact coverage probability for the special case of $R_0 =0$.)}Different path loss exponents are demonstrated: $\alpha = 4$ (left) and $\alpha = 2.5$ (right).}
  \label{fig:topic2_SecureConnectionOutage_C0C5} 
\end{figure}

\begin{figure}[t!]
  \centering
  \subfigure{
    \label{fig:topic2_AverageSecureRate:a} 
    \includegraphics[height=0.273\textheight, bb = 110 263 478 575, clip = true]{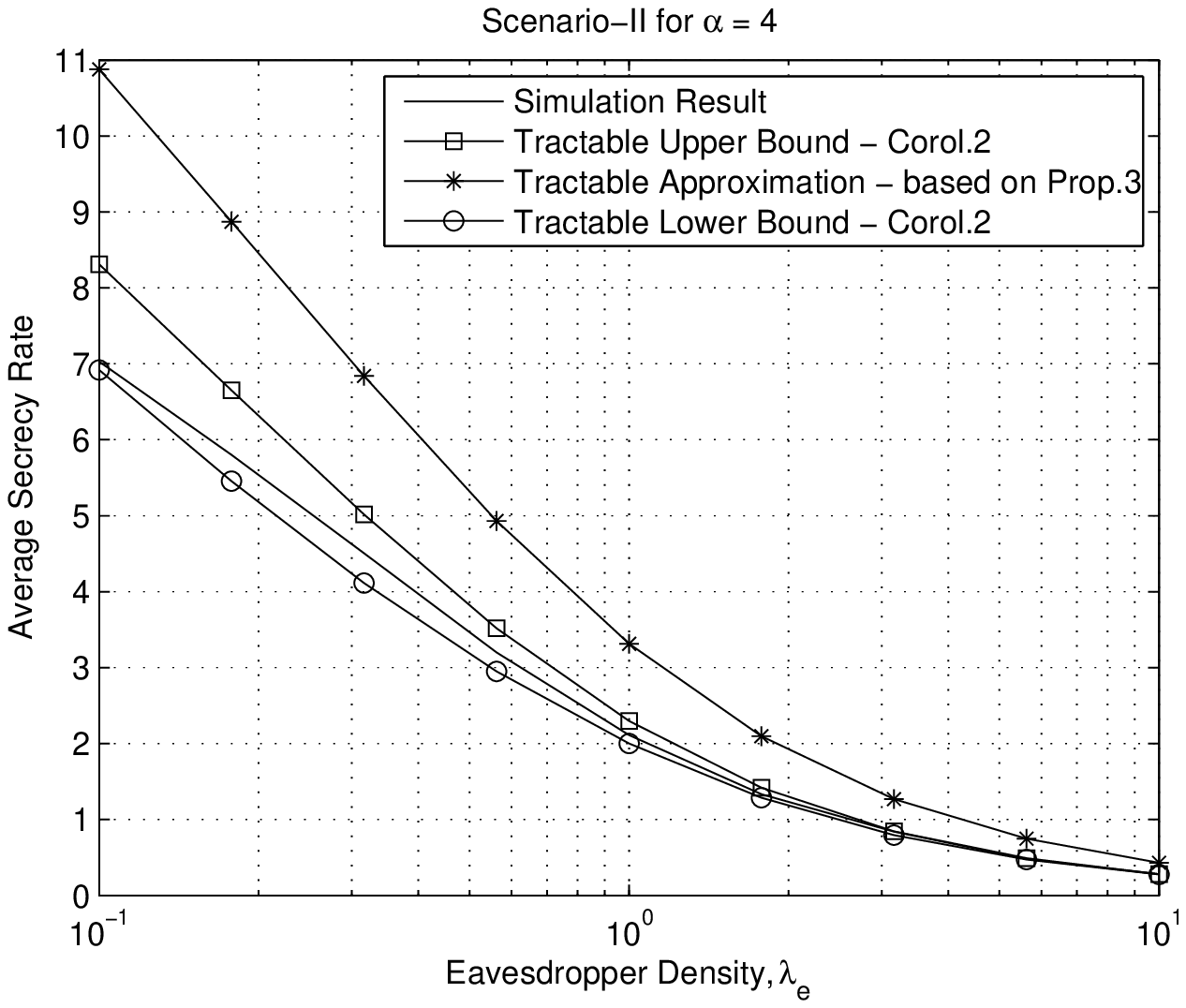}}
  \subfigure{
    \label{fig:topic2_AverageSecureRate:b} 
    \includegraphics[height=0.273\textheight, bb = 110 263 478 575, clip = true]{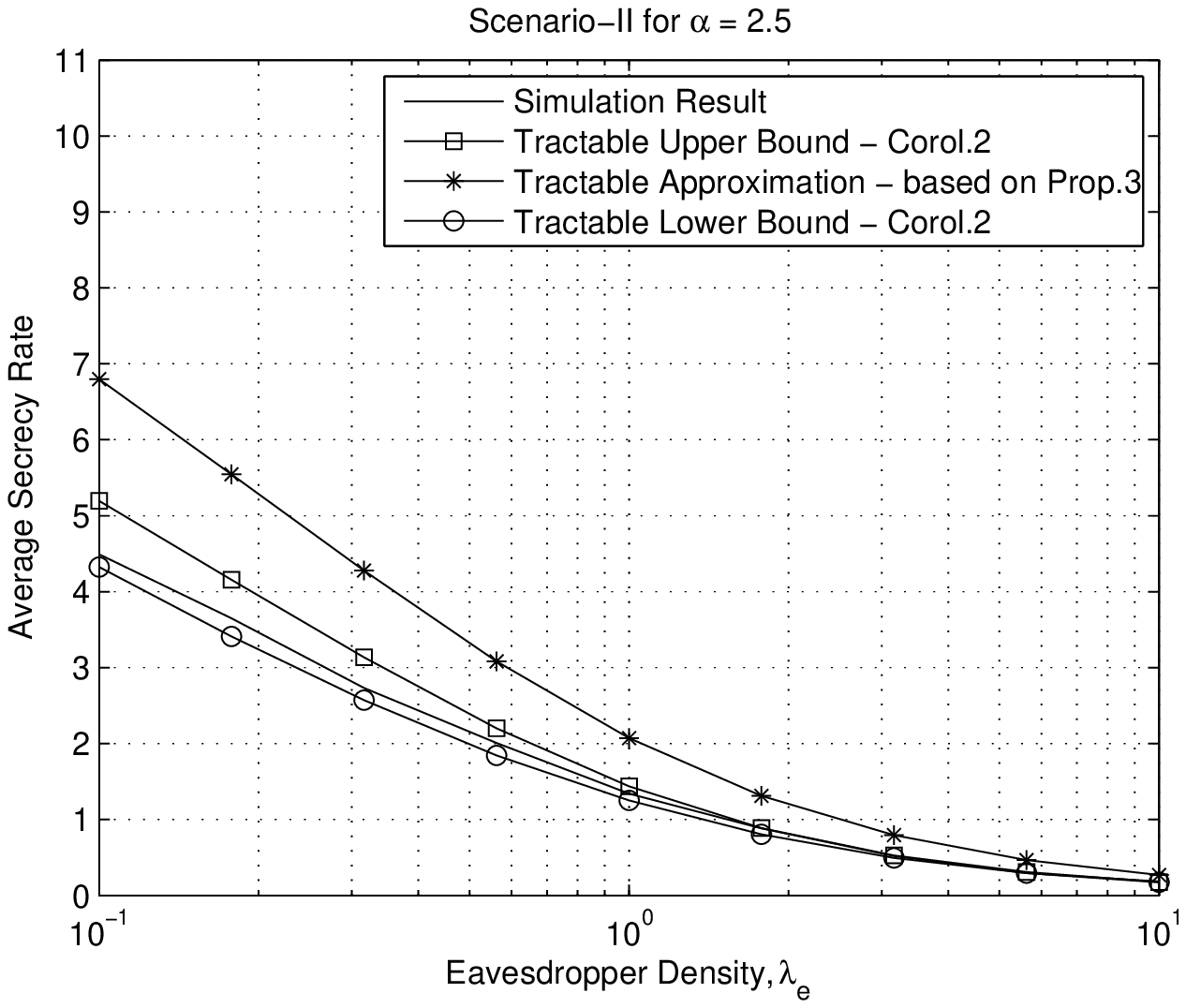}}
  \caption{The average secrecy rate achievable versus the eavesdropper density $\lambda_e$ for
           Scenario-II (full location information; optimal BS to serve).
           Simulation and tractable analytical results are shown for different path loss exponents:
            $\alpha = 4$ (left) and $\alpha = 2.5$ (right).}
  \label{fig:topic2_AverageSecureRate} 
\end{figure}

Fig.~\ref{fig:topic2_SecureConnectionOutage_C0C5} and Fig.~\ref{fig:topic2_AverageSecureRate} demonstrate the results of Scenario-II, the optimal case where all mobile users' location and identity information is completely known and the optimal BS is chosen to maximize the achievable secrecy rate. Fig.~\ref{fig:topic2_SecureConnectionOutage_C0C5} shows the typical user's secure link coverage probability with the threshold $R_0 = 0$ or $R_0 = 5$ to claim outage. Note that the upper bound in Proposition~\ref{prop:FullLocInfo_BestBS_2nd_UpperBound} converges to the exact coverage probability in the special case of $R_0 = 0$, which can be observed from the fact that the curves representing the approximation (\ref{eqn:Cs_upperCCDF_2nd_approx_scn-II}) based on Proposition~\ref{prop:FullLocInfo_BestBS_2nd_UpperBound} match the simulated results in Fig.~\ref{fig:topic2_SecureConnectionOutage_C0C5}. However, this approximation is not precise for large values of $R_0$, e.g., $R_0 = 5$ and the analytical reason for this inaccuracy is explained in remark after Proposition~\ref{prop:FullLocInfo_BestBS_2nd_UpperBound}. On the other hand, the lower bound and the upper bound in Proposition~\ref{prop:FullLocInfo_BestBS} tend to give more accurate approximations of the exact secrecy coverage probability for large values of $R_0$, which can be regarded as a complementary property to offset the limitation of the upper bound in Proposition~\ref{prop:FullLocInfo_BestBS_2nd_UpperBound} mentioned above. From the results shown in Fig.~\ref{fig:topic2_AverageSecureRate}, the tractable upper and lower bounds of the achievable secrecy rates in Corollary~\ref{corol:FullLocInfo_BestBS} are also reasonably accurate. Furthermore, the approximations for the average secrecy rates achievable based on Proposition~\ref{prop:FullLocInfo_BestBS_2nd_UpperBound} are also demonstrated in Fig.~\ref{fig:topic2_AverageSecureRate} and turn out to be inaccurate due to Proposition~\ref{prop:FullLocInfo_BestBS_2nd_UpperBound}'s imprecise estimate for large $R_0$. The achievable secrecy rate given in Scenario-II provides the maximum value over all the scenarios considered in this paper.

By comparing Fig.~\ref{fig:topic1_AverageSecureRate} with Fig.~\ref{fig:topic2_AverageSecureRate}, it can be noted that picking the nearest BS to serve can achieve a secrecy rate nearly as much as the optimal value. For example, the secrecy rate in Scenario-I is approximately 1.9 for $\alpha = 4$ and the eavesdroppers' density $\lambda_e = 1$, compared with around 2.1 for the optimal case in Scenario-II. In other words, there is only marginal benefits from optimally choosing the serving BS instead of simply picking the nearest BS to serve.

\begin{figure}[t!]
  \centering
  \includegraphics[width=0.64\textwidth, bb = 110 260 478 575, clip = true]{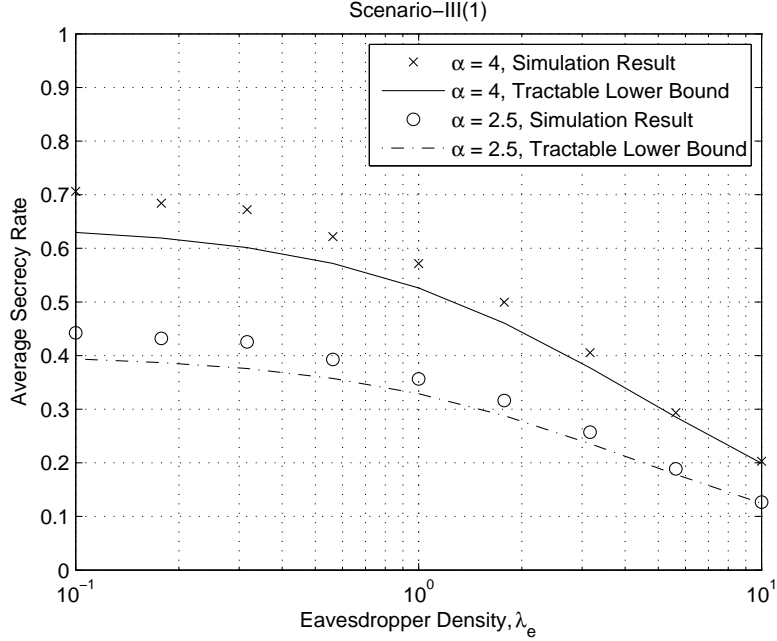}
  \caption{The average secrecy rate achievable versus the eavesdropper density $\lambda_e$ for
           Scenario-III(1) (no location information exchange; nearest BS to serve).
           Simulation results and tractable lower bounds are shown for different path loss exponents $\alpha$.}
  \label{fig:topic4_AverageSecureRate} 
\end{figure}

Fig.~\ref{fig:topic4_AverageSecureRate} shows the average secrecy rate achievable for Scenario-III(1), where no location and identity information exchange is allowed and only intracell users' location information is known to the serving BS. Due to the shrinkage of the region where location information is available, the secrecy performance is significantly degraded compared with the counterpart in Fig.~\ref{fig:topic1_AverageSecureRate}. For example, the average secrecy rate achievable is around $0.57$ for $\alpha = 4$ and $\lambda_e = 1$, whereas the corresponding value can reach around $1.9$ for Scenario-I. We also observe a relatively slow drop in the average secrecy rate achievable as $\lambda_e$ changes from $0.1$ toward $1$, due to its weak dependence on the density of eavesdroppers in this range of $\lambda_e$, which suggests that the lack of location information outside the serving BS's cell becomes the main restrictive factor in determining the secrecy performance. On the other hand, as $\lambda_e$ increases from 1 to $10$, the average secrecy rate achievable accelerates to drop since the eavesdropper density is more influential. It can be shown that the tractable lower bound in (\ref{eqn:Cs_lowerAvg_scn-IIIA}) captures the general trend of the curves and can be used as a tool to make a precise estimate.

\begin{figure}[t!]
  \centering
  \subfigure{
    \label{fig:topic5_AverageSecureRate:a} 
    \includegraphics[height=0.275\textheight, bb = 110 265 478 575, clip = true]{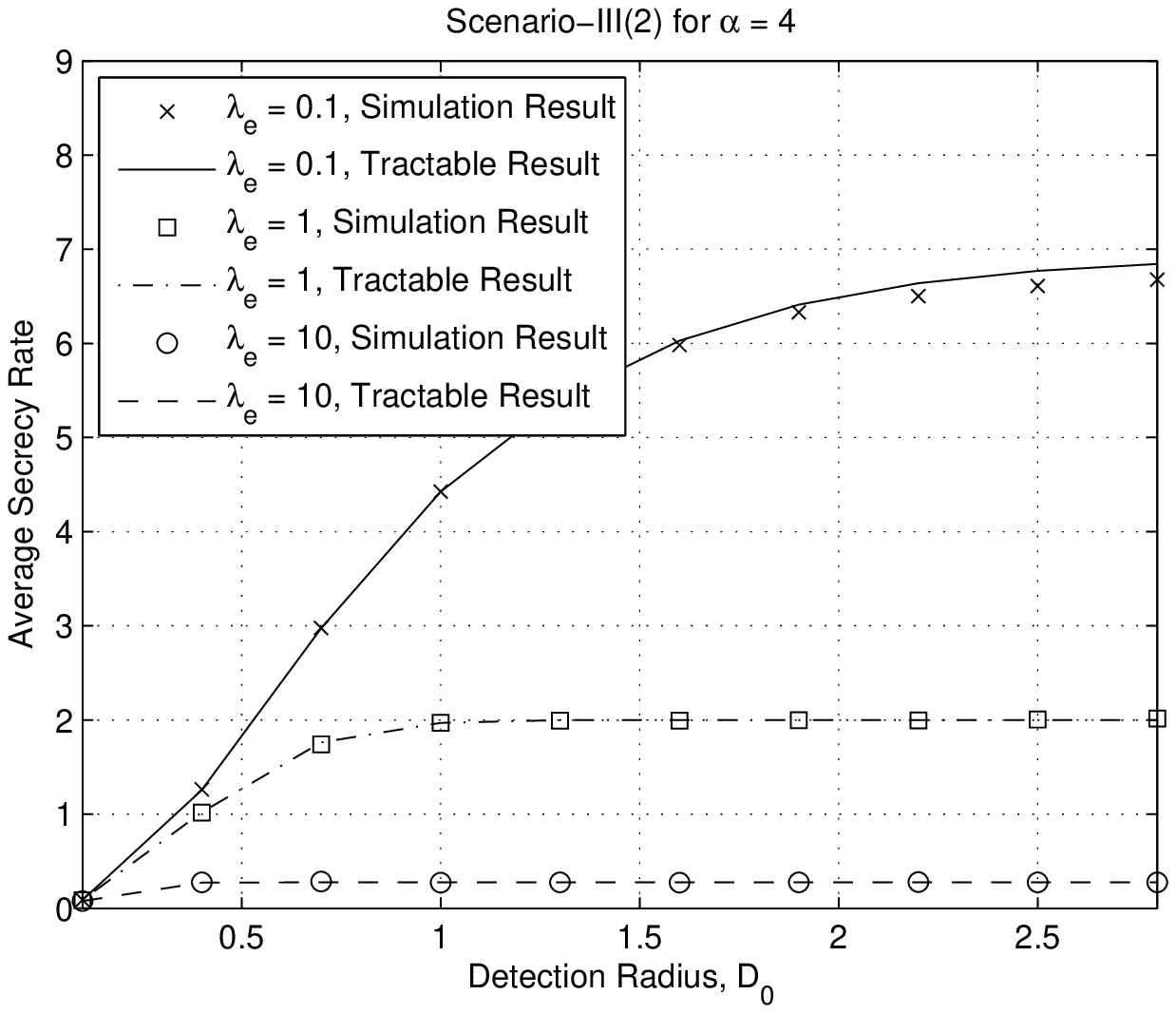}}
  \hspace{-0.03\textwidth}
  \subfigure{
    \label{fig:topic5_AverageSecureRate:b} 
    \includegraphics[height=0.275\textheight, bb = 110 265 478 575, clip = true]{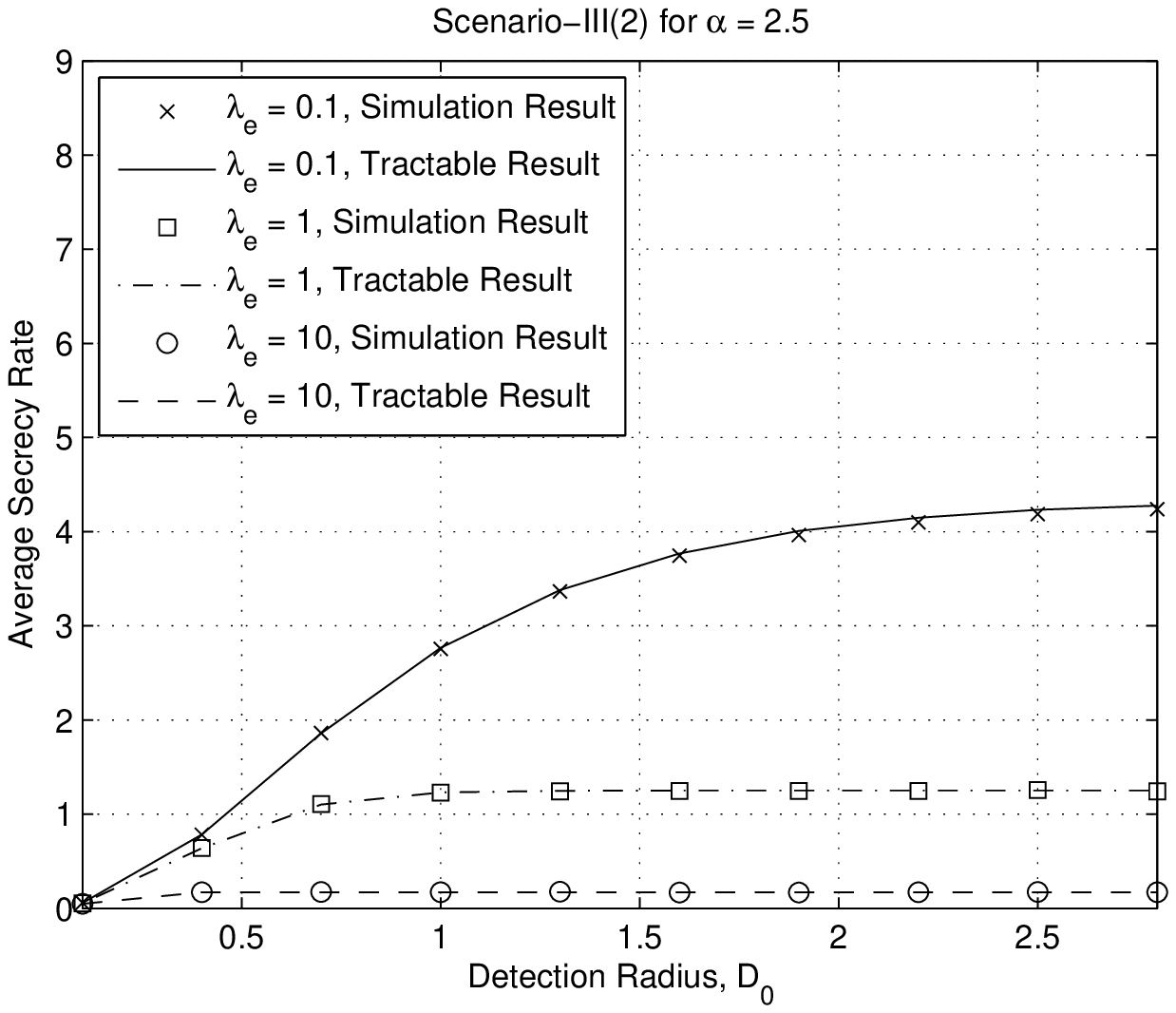}}
  \caption{The average secrecy rate achievable versus the detection radius $D_0$ for
           Scenario-III(2) (location information for users with distances less than $D_0$; nearest BS to serve).
           Simulation and tractable analytical results are shown for different eavesdropper densities $\lambda_{e}$ and different path loss exponents:
            $\alpha = 4$ (left) and $\alpha = 2.5$ (right).}
  \label{fig:topic5_AverageSecureRate} 
\end{figure}

Furthermore, by presenting the average secrecy rate achievable versus the detection radius $D_0$ in Fig.~\ref{fig:topic5_AverageSecureRate}, we can see the importance of eavesdroppers' location information on the secrecy performance. In case of relatively small values of $D_0$, any increase of the detection radius brings remarkable benefit to the achievable secrecy rate. On the other hand, in case of large $D_0$, any further increase in the detection radius does not substantially impact the secrecy rate, since the eavesdropper that limits the secrecy performance is usually located not too far away from the serving BS and its distance is likely to be smaller than $D_0$ when $D_0$ is sufficiently large. Take the curve with $\alpha = 4$ and $\lambda_e = 0.1$ for instance, the secrecy performance improves significantly as $D_0$ is increased up to $2$, and any further increase from $D_0 = 2$ has a limited effect. This performance trend over the range of detection radius can be utilized to appropriately choose the number of neighboring BSs for the information exchange in order to achieve a good secrecy performance whilst taking the implementation cost of such information exchange into consideration. It should be noticed that the slight mismatches between simulation and tractable results in Fig.~\ref{fig:topic1_AverageSecureRate} and Fig.~\ref{fig:topic5_AverageSecureRate} come from the high SNR assumption used in our analysis, and become almost invisible at $\mathrm{SNR} = 30 \mathrm{dB}$ (plots omitted for brevity).

Another fact clearly shown from Fig.~\ref{fig:topic2_AverageSecureRate}-\ref{fig:topic5_AverageSecureRate} is that better performance can be obtained for larger values of path loss exponent $\alpha$, e.g., the average secrecy rate achievable is higher for $\alpha = 4$ than the counterpart for $\alpha = 2.5$. This is because the resultant larger path loss from larger $\alpha$ indicates worse signal condition to both the eavesdroppers and the typical user, whereas the former effect turns out to be more influential on the secrecy performance.

\section{Conclusion}\label{sec:Conclusion}
In this work, we studied the secrecy performance of cellular networks considering cell association and information exchange between BSs potentially provided by the carrier-operated high-speed backhaul and core-networks. Using the stochastic geometry modeling of cellular networks, tractable results to characterize the secrecy rate were obtained under different assumptions on the cell association and location information exchange between BSs. The simulation results validate the tractable expressions and approximations. From the analysis in this paper, we identified the unique feature for secure transmissions that the optimal BS is often not the nearest BS. Nevertheless, our result shows that keeping the nearest BS to be used for secure transmissions still achieves near-optimal secrecy performance. \ignore{Furthermore, by studying the performance difference between the two extremes, i.e., location information fully exchanged and no location information exchange, we reach the conclusion that the location information plays a crucial role in determining the average secrecy rate achievable at the typical user.}We also considered the exchange of eavesdropper's location information between BSs and studied its impact on the secrecy rate performance. Our finding is that it is usually sufficient to allow a small number of neighboring BSs to exchange the location information for achieving close to maximum secrecy rate. Specifically, our analytical result provides network designers practical guidelines to decide the necessary information exchange range, i.e., how many nearby BSs should participate in the information exchange for achieving a certain level of secrecy performance.

The result in this work applies to scenarios where a carefully planned frequency reuse pattern is assumed, and the serving BS can occupy some resource blocks exclusively in a relatively large region. In future cellular networks, however, interference will become an important factor. Since the channel conditions of both legitimate users and eavesdroppers will be degraded by introducing interference, the impact of the co-channel interference on the secrecy performance of large-scale cellular network is still unknown. Another limitation is that the BS cooperation considered in this paper is confined to cell association and location information exchange. Coordinated multipoint (CoMP) transmission, as an emerging BS cooperation technique in future cellular networks, can be potentially utilized, and its benefit on the secrecy performance is an interesting problem to investigate.


%







\bibliographystyle{IEEEtran}
\bibliography{IEEEabrv,Bib_Database}

\begin{thebibliography}{10}
\providecommand{\url}[1]{#1}
\csname url@rmstyle\endcsname
\providecommand{\newblock}{\relax}
\providecommand{\bibinfo}[2]{#2}
\providecommand\BIBentrySTDinterwordspacing{\spaceskip=0pt\relax}
\providecommand\BIBentryALTinterwordstretchfactor{4}
\providecommand\BIBentryALTinterwordspacing{\spaceskip=\fontdimen2\font plus
\BIBentryALTinterwordstretchfactor\fontdimen3\font minus
  \fontdimen4\font\relax}
\providecommand\BIBforeignlanguage[2]{{%
\expandafter\ifx\csname l@#1\endcsname\relax
\typeout{** WARNING: IEEEtran.bst: No hyphenation pattern has been}%
\typeout{** loaded for the language `#1'. Using the pattern for}%
\typeout{** the default language instead.}%
\else
\language=\csname l@#1\endcsname
\fi
#2}}

\bibitem{3GPP_TS33.102}
{3GPP Tech. Spec. 33.102 V8.6.0}, ``{3G security; Security architecture}.''

\bibitem{San09MCOM}
C.~B. Sankaran, ``Network access security in next-generation {3GPP} systems:
  {A} tutorial,'' \emph{{IEEE} Commun. Mag.}, vol.~47, no.~2, pp. 84--91, Feb.
  2009.

\bibitem{Wyn75JBST}
A.~D. Wyner, ``The wire-tap channel,'' \emph{Bell System Technical Journal},
  vol.~54, no.~8, pp. 1355--1387, Oct. 1975.

\bibitem{CsiKor78JIT}
I.~Csisz{\'a}r and J.~K{\"o}rner, ``Broadcast channels with confidential
  messages,'' \emph{{IEEE} Trans. Inform. Theory}, vol.~24, no.~3, pp.
  339--348, May 1978.

\bibitem{Hae08ISIT}
M.~Haenggi, ``The secrecy graph and some of its properties,'' in \emph{Proc.
  {IEEE} Int'l Symp. on Information Theory (ISIT'08)}, Toronto, Canada, July
  2008, pp. 539--543.

\bibitem{PinBar12TIFS1}
P.~C. Pinto, J.~Barros, and M.~Z. Win, ``Secure communication in stochastic
  wireless networks - {Part I}: Connectivity,'' \emph{{IEEE} Trans. Inform.
  Forensics and Security}, vol.~7, no.~1, pp. 125--138, Feb. 2012.

\bibitem{GoeAgg10ISIT}
S.~Goel, V.~Aggarwal, A.~Yener, and A.~R. Calderbank, ``Modeling location
  uncertainty for eavesdroppers: A secrecy graph approach,'' in \emph{Proc.
  {IEEE} Int'l Symp. on Information Theory (ISIT'10)}, Austin, USA, June 2010,
  pp. 2627--2631.

\bibitem{PinWin10ISITA}
P.~C. Pinto and M.~Z. Win, ``Continuum percolation in the intrinsically secure
  communications graph,'' in \emph{Proc. 2010 Int'l Symp. on Information Theory
  and its Applications (ISITA'10)}, Taichung, Taiwan, Oct. 2010, pp. 349--354.

\bibitem{ZhoGan11JWCOM1}
X.~Zhou, R.~K. Ganti, and J.~G. Andrews, ``Secure wireless network connectivity
  with multi-antenna transmission,'' \emph{{IEEE} Trans. Wireless Commun.},
  vol.~10, no.~2, pp. 425--430, Feb. 2011.

\bibitem{PinBar12TIFS2}
P.~C. Pinto, J.~Barros, and M.~Z. Win, ``Secure communication in stochastic
  wireless networks - {Part II}: Maximum rate and collusion,'' \emph{{IEEE}
  Trans. Inform. Forensics and Security}, vol.~7, no.~1, pp. 139--147, Feb.
  2012.

\bibitem{KoyKok12JIT}
O.~O. Koyluoglu, C.~E. Koksal, and H.~E. Gamal, ``On secrecy capacity scaling
  in wireless networks,'' \emph{{IEEE} Trans. Inform. Theory}, vol.~58, no.~5,
  pp. 3000--3015, May 2012.

\bibitem{LiaPoo09ISIT}
Y.~Liang, H.~Poor, and L.~Ying, ``Secrecy throughput of {MANETs} with malicious
  nodes,'' in \emph{Proc. {IEEE} Int'l Symp. on Information Theory (ISIT'09)},
  Seoul, Korea, June 2009, pp. 1189--1193.

\bibitem{CapGoe12Infocom}
C.~Capar, D.~Goeckel, B.~Liu, and D.~Towsley, ``Secret communication in large
  wireless networks without eavesdropper location information,'' in \emph{Proc.
  31st Annual {IEEE} Int'l Conf. on Computer Commun. (IEEE INFOCOM'12)},
  Orlando, USA, Mar. 2012, pp. 1152--1160.

\bibitem{ZhoGan11JWCOM2}
X.~Zhou, R.~K. Ganti, J.~G. Andrews, and A.~Hj{\o}rungnes, ``On the throughput
  cost of physical layer security in decentralized wireless networks,''
  \emph{{IEEE} Trans. Wireless Commun.}, vol.~10, no.~8, pp. 2764--2775, Aug.
  2011.

\bibitem{ZhoGan11Allerton}
------, ``Secrecy transmission capacity of decentralized wireless networks,''
  in \emph{Proc. 49th Annual Allerton Conference on Communication, Control, and
  Computing (Allerton'11)}, Monticello, USA, Sept. 2011, pp. 1726--1732.

\bibitem{StoKen95Book}
D.~Stoyan, W.~S. Kendall, and J.~Mecke, \emph{Stochastic Geometry and its
  Applications}, 2nd~ed.\hskip 1em plus 0.5em minus 0.4em\relax New York, NY:
  John Wiley \& Sons Ltd., 1995.

\bibitem{BacBla09Book}
F.~Baccelli and B.~B{\l}aszczyszyn, \emph{{Stochastic Geometry and Wireless
  Networks, Volume I: Theory}}, 1st~ed.\hskip 1em plus 0.5em minus 0.4em\relax
  Hanover, MA: Now Publishers Inc., 2009.

\bibitem{Bro00JSAC}
T.~X. Brown, ``Cellular performance bounds via shotgun cellular systems,''
  \emph{{IEEE} J. Select. Areas Commun.}, vol.~18, no.~11, pp. 2443--2455, Nov.
  2000.

\bibitem{YanPet03TSP}
X.~Yang and A.~P. Petropulu, ``Co-channel interference modeling and analysis in
  a {Poisson} field of interferers in wireless communications,'' \emph{{IEEE}
  Trans. Signal Processing}, vol.~51, no.~1, pp. 64--76, Jan. 2003.

\bibitem{Hae08TIT}
M.~Haenggi, ``A geometric interpretation of fading in wireless networks:
  {Theory} and applications,'' \emph{{IEEE} Trans. Inform. Theory}, vol.~54,
  no.~12, pp. 5500--5510, Dec. 2008.

\bibitem{AndBac11JCOM}
J.~G. Andrews, F.~Baccelli, and R.~K. Ganti, ``A tractable approach to coverage
  and rate in cellular networks,'' \emph{{IEEE} Trans. Commun.}, vol.~59,
  no.~11, pp. 3122--3134, Nov. 2011.

\bibitem{DhiGan12JSAC}
H.~S. Dhillon, R.~K. Ganti, F.~Baccelli, and J.~G. Andrews, ``Modeling and
  analysis of {K}-tier downlink heterogeneous cellular networks,'' \emph{{IEEE}
  J. Select. Areas Commun.}, vol.~30, no.~3, pp. 550--560, Apr. 2012.

\bibitem{WanQue12JSAC}
W.~C. Cheung, T.~Q. Quek, and M.~Kountouris, ``Throughput optimization,
  spectrum allocation, and access control in two-tier femtocell networks,''
  \emph{{IEEE} J. Select. Areas Commun.}, vol.~30, no.~3, pp. 561--574, Apr.
  2012.

\bibitem{CheNgu12VTC}
C.~S. Chen, V.~M. Nguyen, and L.~Thomas, ``On small cell network deployment: A
  comparative study of random and grid topologies,'' in \emph{Proc. {IEEE} 76th
  Vehic. Tech. Conf. (VTC'12-Fall)}, Qu\'{e}bec City, Canada, Sept. 2012, pp.
  1--5.

\bibitem{YuKim11arXiv}
\BIBentryALTinterwordspacing
S.~M. Yu and S.-L. Kim. Downlink capacity and base station density in cellular
  networks. [Online]. Available: \url{http://arxiv.org/abs/1109.2992}
\BIBentrySTDinterwordspacing

\bibitem{SarHae10IM}
A.~Sarkar and M.~Haenggi, ``{Secrecy Coverage},'' \emph{Internet Mathematics},
  2012, accepted. Available at \url{http://www.nd.edu/~mhaenggi/pubs/im12.pdf}.

\bibitem{BloBar08JIT}
M.~Bloch, J.~Barros, M.~R.~D. Rodrigues, and S.~W. McLaughlin, ``Wireless
  information-theoretic security,'' \emph{{IEEE} Trans. Inform. Theory},
  vol.~54, no.~6, pp. 2515--2534, June 2008.

\bibitem{Hae05TIT}
M.~Haenggi, ``On distances in uniformly random networks,'' \emph{{IEEE} Trans.
  Inform. Theory}, vol.~51, no.~10, pp. 3584--3586, Oct. 2005.

\bibitem{JefDai08HandBook}
A.~Jeffrey and H.-H. Dai, \emph{Handbook of mathematical formulas and
  integrals}, 4th~ed.\hskip 1em plus 0.5em minus 0.4em\relax Burlington, MA:
  Academic Press, 2008.

\bibitem{OkaBoo92Book}
A.~Okabe, B.~Boots, and K.~Sugihara, \emph{Spatial Tessellations: Concepts and
  Applications of {Voronoi} Diagrams}, 1st~ed.\hskip 1em plus 0.5em minus
  0.4em\relax West Sussex, England: John Wiley \& Sons Ltd., 1992.

\bibitem{HinMil80Math}
A.~L. Hinde and R.~E. Miles, ``{Monte Carlo} estimates of the distributions of
  the random polygons of the {Voronoi} tessellation with respect to a {Poisson}
  process,'' \emph{Journal of Statistical Computation and Simulation}, vol.~10,
  no. 3-4, pp. 205--223, 1980.

\bibitem{WeaKer86Math}
D.~Weaire, J.~P. Kermode, and J.~Wejchert, ``On the distribution of cell areas
  in a {Voronoi} network,'' \emph{Philosophical Magazine Part B}, vol.~53,
  no.~5, pp. L101--L105, 1986.

\bibitem{AbrSte65HandBook}
M.~Abramowitz and I.~Stegun, \emph{Handbook of mathematical functions: with
  formulas, graphs, and mathematical tables}, 1st~ed.\hskip 1em plus 0.5em
  minus 0.4em\relax Mineola, NY: Dover Publications, 1965.

\end{thebibliography}

\newpage

\end{document}